%% file: nika5.tex
\documentclass[twocolumn,traditabstract]{aa}

\usepackage[english]{babel}
\usepackage{graphicx,amsmath}
\usepackage{epstopdf}
\usepackage{epsf,color}
\usepackage[mathscr]{eucal}
\usepackage{amsmath}
\usepackage{amssymb,amsfonts}
\usepackage{natbib}
\usepackage[breaklinks, colorlinks, citecolor=blue, linkcolor=MyBlue, urlcolor=RoyalPurple, colorlinks=true, linkcolor=blue, debug, baseurl=' ']{hyperref}
\bibpunct{(}{)}{;}{a}{}{,} 

\def\simlt{\lower.5ex\hbox{$\; \buildrel < \over \sim \;$}}
\def\simgt{\lower.5ex\hbox{$\; \buildrel > \over \sim \;$}}
\def\NIKA{\textit{NIKA}}

\def\NIKAii{\textit{NIKA2}}
\def\Skydip{\textit{Skydip}}
\begin{document}

\title{Performance and calibration of the NIKA camera at the IRAM~30~m telescope}

\author{
A.~Catalano$^1$,
M.~Calvo$^2$,
N.~Ponthieu$^3$,
R.~Adam$^1$,
A.~Adane$^4$,
P.~Ade$^5$,
P.~Andr\'e$^6$,
A.~Beelen$^7$,
B.~Belier$^8$,
A.~Beno\^it$^2$,
A.~Bideaud$^5$,
N.~Billot$^9$,
N.~Boudou$^2$,
O.~Bourrion$^1$,
G.~Coiffard$^4$,
B.~Comis$^1$,
A.~D'Addabbo$^{2,11}$,
F.-X.~D\'esert$^3$,
S.~Doyle$^5$,
J.~Goupy$^2$,
C.~Kramer$^9$,
S.~Leclercq$^4$,
J.~F.~Mac\'ias-P\'erez$^1$,
J.~Martino$^7$,
P.~Mauskopf$^{5,12}$,
F.~Mayet$^1$,
A.~Monfardini$^2$,
F.~Pajot$^7$,
E.~Pascale$^5$,
L. Perotto$^1$,
V.~Rev\'eret$^6$,
L.~Rodriguez$^6$,
G.~Savini$^{10}$,
K.~Schuster$^4$,
A.~Sievers$^9$,
C.~Tucker$^5$,
R.~Zylka$^4$}

\institute{
Laboratoire de Physique Subatomique et de Cosmologie,
Universit\'e Grenoble Alpes, CNRS/IN2P3,
  53, rue des Martyrs, Grenoble, France
\and
Institut N\'eel, CNRS and Universit\'e de Grenoble, France
\and
Institut de Plan\'etologie et d'Astrophysique de Grenoble (IPAG), CNRS and Universit\'e de
Grenoble, France
\and
Institut de Radio Astronomie Millim\'etrique (IRAM), Grenoble, France
\and
Astronomy Instrumentation Group, University of Cardiff, UK
\and
Laboratoire AIM, CEA/IRFU, CNRS/INSU, Université Paris Diderot, CEA-Saclay, 91191 Gif-Sur-Yvette, France 
\and
Institut d'Astrophysique Spatiale (IAS), CNRS and Universit\'e Paris Sud, Orsay, France
\and
Institut d'Electronique Fondamentale (IEF), Universit\'e Paris Sud, Orsay, France
\and
Institut de Radio Astronomie Millim\'etrique (IRAM), Granada, Spain
\and
University College London, Department 
of Physics and Astronomy, Gower Street, London WC1E 6BT, UK
\and
Universit\`a \textit{Sapienza} di Roma, Italy
\and
SESE and Physics, Arizona State University, Tempe, AZ, USA
}

\date{\today}

\input{00_abstract}


\authorrunning{NIKA coll.}
\titlerunning{NIKA camera calibration and performance}

\maketitle

\newpage

\section{Introduction} 
\label{introduction}
\input{01_introduction}

\section{The \NIKA\ Instrument}
\label{instr}
\input{02_instrument}

\section{Readout optimization procedure}
\label{tuning}
\input{03_tuning}

\section{Field-of-view (FOV) properties and main beam characterization}
\label{focal}
\input{04_focal}

\section{Data reduction}
\label{dataprocessing}
\input{05_toiproc}

\section{Photometric calibration} 
\label{abs_cal}
\input{06_photcal}

\section{Noise equivalent flux density (NEFD)}
\label{se:nefd}

\input{07_nefd}

\section{NIKA observations}
\label{obs}
\input{08_skyobs}

\section{Conclusion and perspectives} 
\label{conclusion}
\input{09_conclusion}

\begin{acknowledgements}
We would like to thank the IRAM staff for their support during the campaign. 
This work has been partially funded by the Foundation Nanoscience Grenoble, the ANR under the contracts \textit{MKIDS} and \textit{NIKA}. 
This work has been partially supported by the LabEx FOCUS ANR-11-LABX-0013. 
This work has benefited from the support of the European Research Council Advanced Grant ORISTARS under the European Union's Seventh Framework Program (Grant Agreement no. 291294).
The NIKA dilution cryostat was designed and built at the Institut N\'eel. We acknowledge the crucial contribution of the Cryogenics Group and, in particular, Gregory Garde, Henri Rodenas, and Jean Paul Leggeri. R. A. would like to thank the ENIGMASS French LabEx for funding this work. B. C. acknowledges support from the CNES post-doctoral fellowship program.
\end{acknowledgements}

\bibliographystyle{aa}
\bibliography{nika5}

\end{document}

%% file: 00_abstract.tex
\abstract
{The New IRAM KID Array (\NIKA) instrument is a dual-band imaging camera
operating with Kinetic Inductance Detectors (KID) cooled at 100~mK.  \NIKA\
is designed to observe the sky at wavelengths of 1.25 and 2.14~mm from
the IRAM 30~m telescope at Pico Veleta with an estimated resolution of 13\,arcsec
and 18~arcsec, respectively.  This work presents the performance
of the \NIKA\ camera prior to its opening to the astrophysical community as an
IRAM common-user facility in early 2014. \NIKA\ is a test bench for
the final \NIKAii\ instrument to be installed at the end of 2015. The last
\NIKA\ observation campaigns on November 2012 and June 2013 have been used to
evaluate this performance and to improve the control of systematic effects.
We discuss here the dynamical tuning of the readout electronics to optimize the
KID working point with respect to background changes and the new technique of atmospheric absorption correction. These modifications significantly improve the overall linearity, sensitivity, and absolute calibration
performance of \NIKA. This is proved on observations of point-like sources for
which we obtain a best sensitivity (averaged over all valid detectors) of 40 and
14~mJy.s$^{1/2}$ for optimal weather conditions for the 1.25 and 2.14~mm arrays,
respectively.  \NIKA\ observations of well known extended sources (DR21
complex and the Horsehead nebula) are presented.  This performance
makes the \NIKA\ camera a competitive astrophysical instrument.}

%% file: 01_introduction.tex

New challenges in millimeter-wave astronomy require instruments with a combination of high sensitivity, angular resolution, and mapping speed.  These goals demand 
the development of   a new generation of arrays with large number of 
detectors (see for example \cite{2010SPIE.7741E...2B, gismo, 2008ACT, 2008SPIE.7020E..25G}). Current individual detectors, such as high
impedance bolometers \citep{2012MmSAI..83...72T}, are already photon-noise-limited both for space and the ground observations. 

One of the possible technologies suitable for scaling up to larger format arrays is the 
kinetic inductance detector (KID). These detectors are designed to be read out with frequency-division 
multiplexing with a large multiplex factor from a few hundred to several thousand per 
readout chain \cite{2012SPIE.8452E..0OB, Swenson2010}) and are relatively simple
to fabricate (see for example \cite{day2003, 2007stt..conf..170D}).
This technological solution has been selected
for the \NIKA\ project (\cite{NIKA_2010, NIKA_2011}) consisting of
a dual-band millimeter camera for observations at the 30~m Institut de RadioAstronomie Millim\'etrique (IRAM) telescope (Pico Veleta, Spain).
\\ 
The \NIKA\ camera is a 356-pixel instrument
conceived as a resident common user instrument and 
an early technology demonstrator of the competitiveness of KID arrays. \NIKA\ also serves
as preparation for \textit{NIKA2}, a full-scale instrument to be deployed in late 2015 and 
designed to fill the field of view of the IRAM 30~m telescope. 
The goals of the \NIKA\ instrument are to perform simultaneous observations in two bands (1.25\,mm and
2.14\,mm) of milliJansky point sources and to map faint extended
continuum emission with diffraction-limited
resolution and background-limited performance. Achieving these goals 
will enable the \NIKA\ instrument to be competitive in several
astrophysical fields, such as observations from the northern hemisphere of clusters of galaxies detected by PLANCK via the Sunyaev-Zel'dovich (SZ) effect (\cite{2013A&A...550A.128P}), observations of high redshift
galaxies and quasars, detection of early stages of star formation in molecular clouds in our galaxy 
and mapping of dust and free-free emission in nearby galaxies.  In a companion paper, we present the first observation of the thermal SZ effect on
clusters of galaxies \citep{2013arXiv1310.6237A}.  The \NIKA\ camera is now open to the public for observations as of January 2014.

Previous observing campaigns with the \NIKA\ camera have revealed several technical aspects that limited the sensitivity of detectors. \cite{Calvo2013} show how to linearize the KID signal, while another important issue is the working point of the detectors that evolves with time, owing to varying atmospheric conditions,
thus producing a loss in response. A dynamical tuning of the readout
electronics was developed to optimize the KID working point
between two different sky observations. Atmospheric absorption correction is also required, and a dedicated procedure has been developed in order to use the \NIKA\ instrument as a tau-meter.

These improvements were implemented for the last two technical observing
campaigns that took place in November 2012 and June 2013. We report here  the overall
linearity, sensitivity, and absolute calibration of \NIKA\ .

This article is structured as follows: a review of the main characteristics of
the \NIKA\ instrument that focuses on the instrumental improvements with respect
to the previous observing campaigns, is presented in Sects.~\ref{instr} and~\ref{tuning}.
In Section~\ref{focal}, we describe the sky observations made with \NIKA\ that
allowed us to characterize the focal plane. Section~\ref{dataprocessing} provides a short description of the reduction pipeline used to analyse the data. In Section~\ref{abs_cal} we describe the atmospheric absorption
correction method. In Section~\ref{se:nefd}, the final performance
of \NIKA\ in terms of  instrumental noise equivalent flux density (NEFD) is
calculated on point-like sources. Finally in Section~\ref{obs} we {detail the} observations
of point sources and extended sources of the last November 2012 (run5) and June
2013 (run6) campaigns.

%% file: 02_instrument.tex
\begin{figure}[t!]
\begin{center}
\includegraphics[scale=0.25]{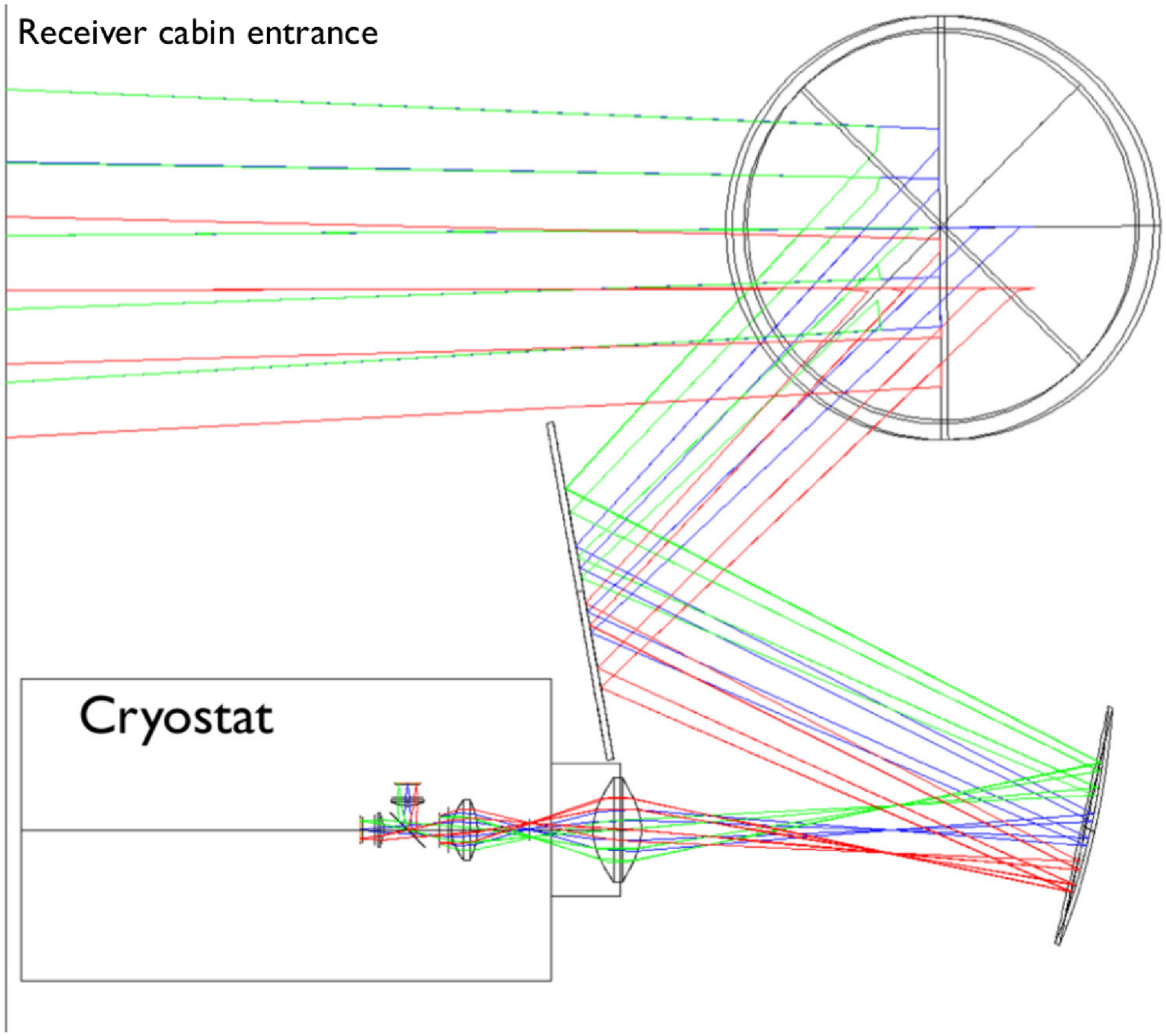}
\includegraphics[scale=0.25]{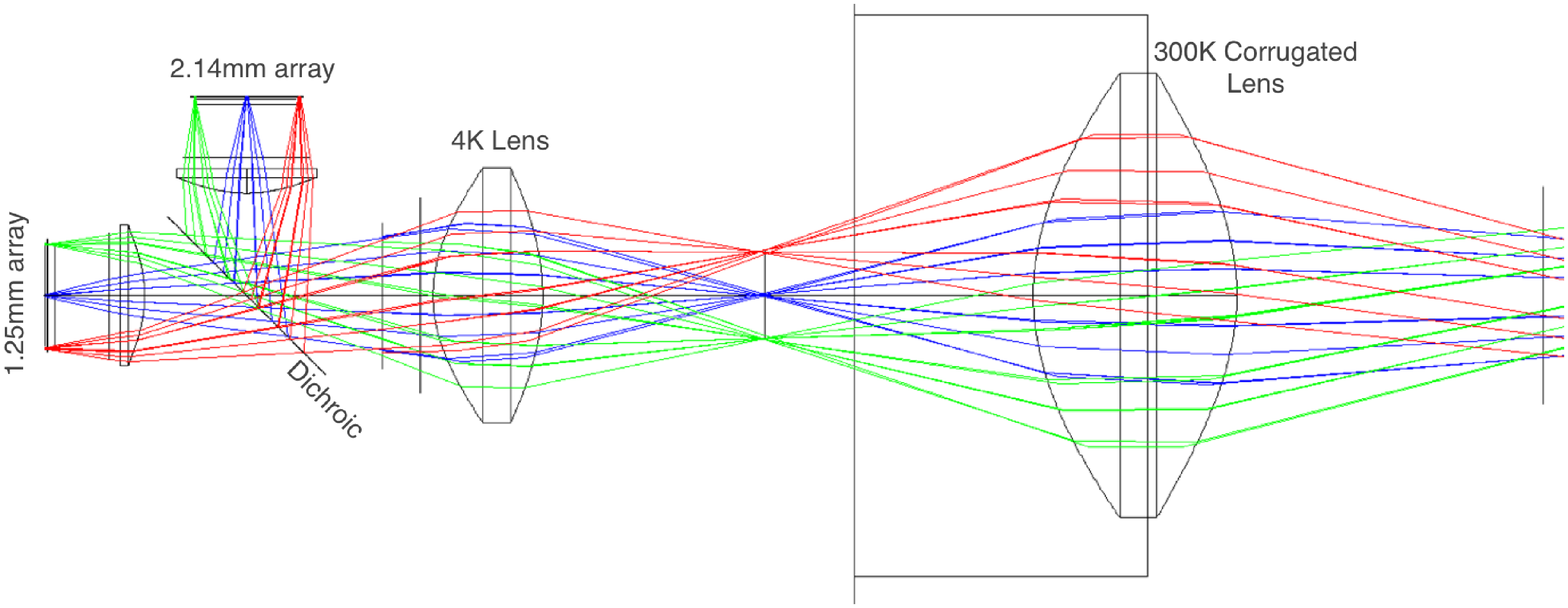}
\end{center}
\caption{Snapshot from the Zemax simulation used to optimize the optical
  system of \NIKA. In the order of photon travel, the \NIKA\ optics consist of a flat mirror at the top of the cryostat, an off-axis biconic-polynomial curved mirror, a 300 K window lens, a field stop, a 4 K lens, an aperture stop, a dichroic, a 100 mK lens, and two band- defining filters in front of the back-illuminated KID arrays. Top panel: ray tracing from the entrance of the
  receiver cabin of the 30~m telescope (which is simulated but not shown in the
  image), view from the elevation axis (which is symbolized by the big circles
  and contains the 2 mirrors of the Nasmyth system). The \NIKA\ cryostat and
  entrance nose are represented by the rectangles. Bottom panel: zoomed plot of the optical system inside the \NIKA\ cryostat. The angular size, the beam efficiency and the overall optical efficiency of the system are presented in Table~\ref{tab:nika_char}.}
\label{fig:optics}
   \end{figure}

Cooling the two KID arrays at about 100~mK is the major requirement for
driving the architecture of the \NIKA\ instrument. This is achieved by a 4~K
cryocooler and a closed-cycle $^3$He - $^4$He dilution.  The optical coupling
between the telescope and the detectors is made by warm aluminum mirrors and
cold refractive optics.  The optics consists of a flat mirror at the
top of the cryostat, an off-axis biconic-polynomial curved mirror, a 300~K
window lens, a field stop, a 4~K lens, an aperture stop, a dichroic, a 100~mK
lens, and two band-defining filters in front of the back-illuminated KID
arrays, which have a backshort that matches the corresponding wavelength.  A view
of the optics of the \NIKA\ instrument and the coupling with the 30~m
telescope is presented in Fig. \ref{fig:optics}.  All the elements presented
in this figure are real optical elements except for the vertical segment
showing the entrance of the receiver cabin and the vertical segments inside
the cryostat (field and pupil diaphragms, filters, dichroic and detector
arrays). 

The throughput of each pixel is calculated using the knowledge of the telescope diameter and equivalent focus, the fraction of unvignetted pupil defined by the cold field stop, and the receiver pixel (KID) size with respect to diffraction pattern. Using $A$ for Area, $\Omega$ for solid angle, $u$ for pixel size in unit of $F \lambda$ ($\lambda$ = wavelength and F= f-number or relative aperture), we have
$$
A\Omega_{pixel} = \frac{A_{M1} (u F \lambda)^2}{f^2_{eq}} = (\pi/4)  (u \lambda)^2 .
$$
where $A_{M1}$ is the effective primary mirror diameter, and $f_{eq}$ the equivalent focal of the telescope.
The effective throughput of the instrument is then $A \Omega_{effective} = N*A\Omega_{pixel}$, where N is the number of valid receiver pixels.

\begin{figure}[t]
\begin{center}
\includegraphics[scale=0.52]{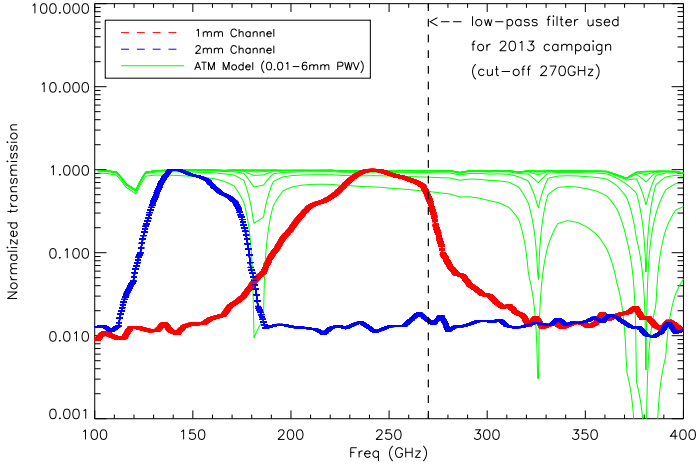}
\end{center}
\caption{\NIKA\ bandpasses.  The bandpass of the 1.25~mm channel (resp. 2.14~mm)
  is shown in red (resp. blue). The bandpasses are averaged over all valid
  pixels, with dispersion (rms) of 2~\% at 1.25~mm and 1~\% at 2.14~mm.  The ATM
  model calculated for different water vapor contents is presented in green.}
\label{fig:bandpass}
   \end{figure}

   The rejection of out-of-band emission from the sky and the telescope is achieved
   by using a series of low-pass metal mesh filters, placed at different
   cryogenic stages in order to minimize the thermal loading on the
   detectors. This determines the shape, width, and position of each of the \NIKA\
   bands.  Spectral characterization of the \NIKA\ bandpass was performed using a
   Martin-Puplett interferometer allowing recovery of the spectral performance
   of each pixel of the two \NIKA\ channels with uncertenties of a few
   percent. In Fig \ref{fig:bandpass} we present the \NIKA\ bandpasses, together
   with the ATM model calculated for different water vapor contents
   (\cite{2001IEEE....49.1683C}). According to the Pardo model, the 1.25~mm
   channel is almost sensitive exclusively to the water vapor emission (water vapor
   secondary line at 183~GHz) in contrast to the 2.14~mm channel, which is slightly sensitive to the roto-vibrational emission line of dioxygen (at 119~GHz). ATM also predicts the 
   contributions of minor species like ozone to the atmospheric absorption. These
   additional contributions are ignored in this paper. Changing the
   precipitable water vapor (pwv) content in the ATM model, we can derive the
   expected sky opacities integrated into the \NIKA\ channels. The ratio between the
   opacities derived for the 1.25~mm and 2.14~mm channels is obtained not only
   according to the pseudo continuum emission of the atmosphere (proportional
   to $\nu^2$ where $\nu$ represents the electromagnetic frequency) but also taking the contribution of the water vapor emission and dioxygen emission into account.  Furthermore, between the 2012 and 2013
   observing campaigns we changed (for technical reasons) the cold optical filter 
   chain by adding a low pass filter with a
   cut-off at 270GHz. Therefore, we expect to have a different ratio in the
   derived sky opacities at 1.25~mm and 2.14~mm channels between the two campaigns. From
   the model, we obtain a ratio between the in-band sky opacities $\tau(2.14~mm)/\tau(1.25~mm)$ of 0.75 and 0.6 for the 2012 and 2013 observing campaigns, respectively.

\begin{table*}
\begin{center}
\begin{tabular}{ccc}
\hline
\hline
 & 1.25~mm channel &  2.14~mm channel  \\
\hline \hline
Pixel size [mm] & 1.6   & 2.3 \\
Dual polarization &  yes & yes \\
Angular size ($F \lambda$) &  0.9 & 0.79 \\
Beam efficiency [\%] &  55 & 70 \\
Detector efficiency [\%] & 80 & 80 \\
Overall optical efficiency [\%] &  30 & 30  \\
Total background [pW] &  45 & 20  \\
\hline \hline
\end{tabular}
\end{center}
\caption{Characteristics of the NIKA instrument coupled to the IRAM telescope. The total background is calculated per pixel with a contribution of the atmosphere derived in good weather conditions. This corresponds to an expected photon-noise level equal to about $5 \cdot 10^{-17} W/\sqrt{Hz}$ for the 2.14mm channel and $9 \cdot 10^{-17} W/\sqrt{Hz}$ for the 1.25mm channel. These parameters have a 50\% precision level.}
\label{tab:nika_char}
\end{table*}

\subsection{Detector array configurations}\label{dac}
The \NIKA\ LEKID detector arrays are made of aluminum and are sensitive to dual polarization. The need to have a symmetric design with a constant filling factor over the whole direct sensitive area drives us to use Hilbert fractal design, which is a well-known geometry for patch antennas (\cite{Roesch2012}). The gap of the Al films has been measured thanks to absorption spectra taken in the lab using a Martin-Pupplet interferometer. The measurements show that below roughly 110~GHz, no radiation is absorbed, meaning that the energy gap in the case of 18nm Al films is approximately 0.2~meV, and the Tc around 1.45~K. The expected film impedance (with respect to the incoming photon able to break Cooper pairs) is actually that of the normal film state. This has been measured to be around 2~$\Omega$/square. This value has been used when designing the detectors to match their effective impedance to that of the incoming radiation (which is absorbed from the substrate side, in a \textit{back-illumination} configuration).

The results reported in this paper cover two different observing campaigns at the telescope, with slightly different properties of the two focal planes:

\begin{figure}[t!]
\begin{center}
\begin{tabular}{c}
\hspace{0.5cm}
\includegraphics[bb = 1 1 170 104,width=6.4cm,clip]{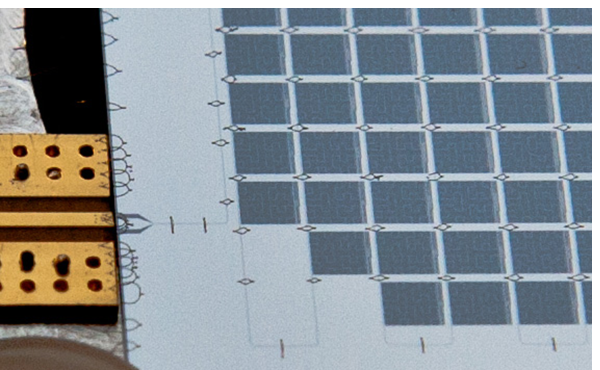}\\
\includegraphics[bb = 2 238 562 598,width=7cm,clip]{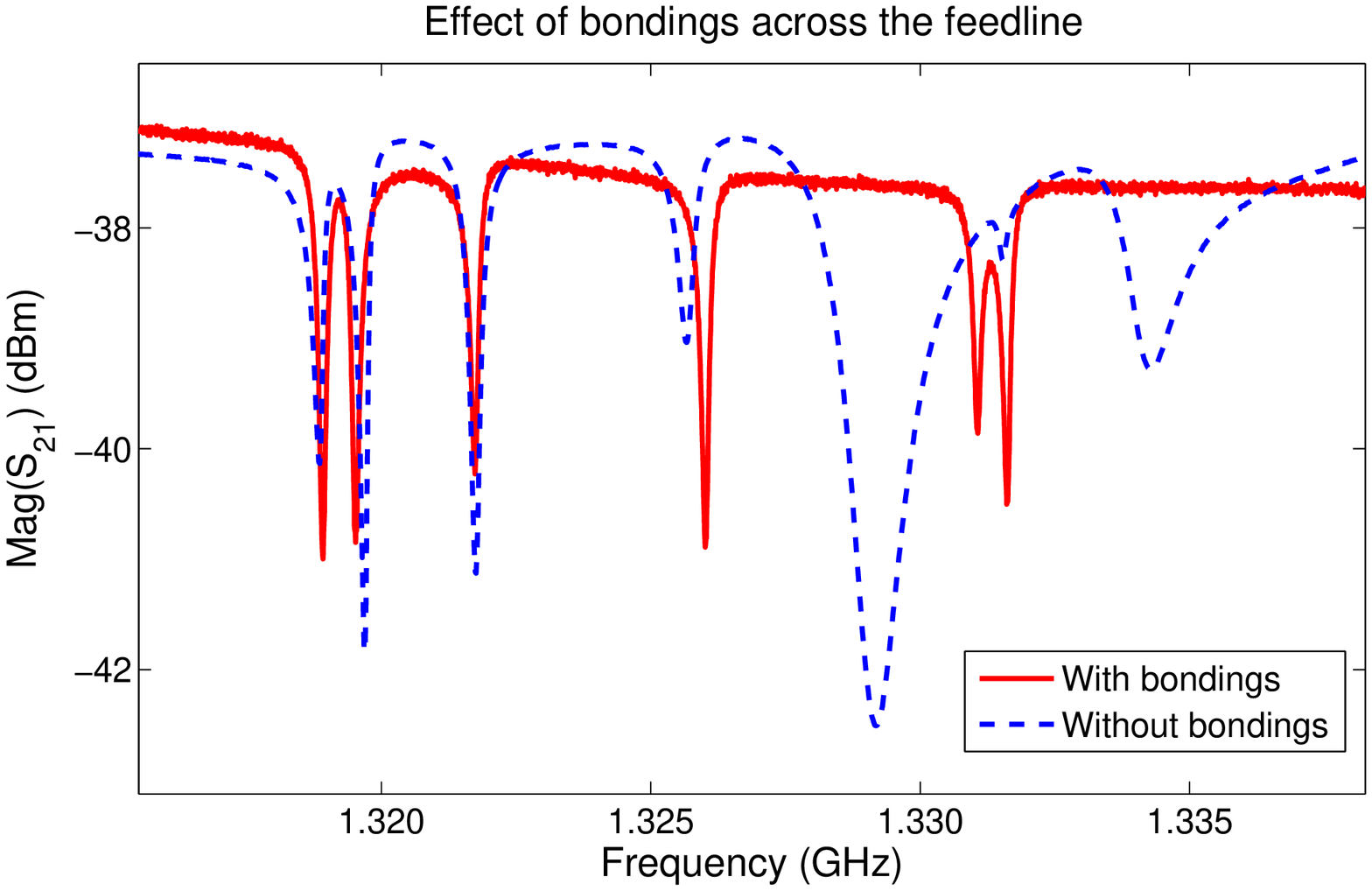}
\end{tabular}
\end{center}
\caption{Image of the bondings added across the feedline (top). As can be seen
  in the frequency sweep carried out before  (blue) and after (red) adding the bondings
  (bottom), the depth of the different resonances has become much more
  uniform, since the coupling of the resonators to the feedline is no longer
  affected by the presence of standing waves supported by the slotline
  modes. Such standing waves were also responsible of the larger dips that are
  observed before adding the bondings, and they disappear afterwards.}
\label{fig:bondings}
\end{figure}

\begin{itemize}
\item{\textbf{Campaign 11/2012}. During this campaign, referred to as Run 5,
    the 2.14~mm array is made of 132 pixels with a 20~nm thick Al film on a
    300~$\mu$m HR silicon substrate. The 1.25~mm array consists of 224~pixels
    with a film of the same thickness but on a 180~$\mu$m substrate. The cold
    amplifier of the 240~GHz channel showed an unexpectedly low saturation
    power, so that we could simultaneously read only eight pixels out with the
    ideal excitation level or, alternatively, 90 pixels but with a lower power
    per tone, resulting in suboptimal performance.  The number of valid
    pixels were 100 and 80 for the 2.14~mm and 1.25~mm channels, respectively.
    The configuration adopted during this observing campaign limited the final sensitivity of the instrument. 
    The channel at 2.14~mm (changed for the 2013 campaign) was limited by the detector noise. For the 1.25~mm channel, 
    the detectors noise was always relatively flat. This means that the intrinsic frequency noise 
    due to random variations of the effective dielectric is negligible compared to the sky noise. 
    The cold amplifier temperature noise was also negligible compared to the sky noise.}
\item{\textbf{Campaign 06/2013}. This campaign is referred to as Run 6. For
    the 2.14~mm channel, we replaced the array with a new one, obtained from a
    thinner Al film (18~nm), in which we optimized the coupling further to the
    feedline. We also added bondings across the feedline to suppress spurious
    slotline modes that were affecting the uniformity of the pixel properties,
    which led to an increase in the number of valid pixels (figure
    \ref{fig:bondings}). The overall geometry and the pitch between pixels
    was left unchanged, as was the readout chain. For the 1.25~mm channel, the
    only intervention was to replace the cold amplifier with a new one
    having a higher power handling. RF filters have also been added on the
    readout chain to avoid the harmonics of the local oscillator from
    reaching the cold amplifier input. Since we suspected that the optical
    load on the 1.25~mm array was preventing it from cooling down
    appropriately, we added an additional low-pass edge filter cutting
    frequencies above 270~GHz, even though this led to the loss of a fraction 
    of the power available in the atmospheric window of
    interest. The number of valid pixels for this campaign was 125 at 2.14~mm
    and 190 at 1.25~mm, for a total of 315 pixels.
   This new configuration was expected to yield a much improved low noise performance, but the bad weather 
   during the observing campaign limited the sensitivity of our arrays. For both channels the dominant noise contribution was due to residual sky noise associated to atmospheric turbulences and residual correlated electronic noise.}
\end{itemize}



For the readout, we used the new NIKEL version 1 electronics, which have
worked flawlessly. The boards, described in detail in \cite{Bourrion2012}, are
capable of generating up to 400 tones each over a 500~MHz bandwidth. This is
achieved by using six separate FPGAs: five of them generate 80 tones each over a
100~MHz band, using five associated DACs. The sixth FPGA acts as a central unit
that combines the signal of the other units, apprioprately shifting and
filtering the different 100~MHz sub-bands to finally cover the whole 500~MHz
available for the frequency comb used to excite the detectors. An analogous,
but reversed, process is then applied to the signal acquired by the ADC of the
board, which is once again split in five different sub-bands treated
separately. Each NIKEL v1 board thus allows us to monitor all the 400 tones
simultaneously with a margin of two bits in the 12 bits ADC dynamics.



Data are acquired at a 23.842~Hz rate, synchronously over the two arrays. Binary
files are provided as output, which contain the raw data, as well as the
resonant frequency of each KID used to set the corresponding excitation tone.

The principal characteristics of the NIKA instrument are presented in Table \ref{tab:nika_char}. 


%% file: 03_tuning.tex

In the standard KID readout scheme, each pixel is excited using a fixed tone at its resonant frequency. The signal transmitted past the detector is compared to a reference copy of the excitation signal to get its in-phase ($I$) and quadrature ($Q$) components. From these it is then necessary to estimate the corresponding shift in resonance frequency $f_0$ of the detector, because this is the physical property directly related to the incoming optical power: $\delta f_0 \propto \delta P_{opt} $ for low values of $\delta P_{opt} $ (\cite{Swenson2010}).
Finding a reliable way to evaluate $f_0(t)$ starting from $I(t)$ and $Q(t)$ represents a challenge, and this is especially true in the case of ground-based experiments, as these have to cope with the effects induced by the variations in atmospheric opacity. To improve the photometric reproducibility, we have developed a measurement process which is described in the next subsection.

\subsection{Modulated Readout}\label{REU}
\label{sec:rfdidq}

For the \NIKA\ detectors an innovative readout technique has been developed, which has been succesfully tested during the 2011 run and adopted for all the following campaigns. The details of this technique are described in \cite{Calvo2013}. Briefly, the underlying idea is to replace the standard excitation of the detectors, which uses a fixed tone, with a new excitation based on two different frequencies. We achieve this by modulating the local oscillator signal between two values, separated by $\delta f_{LO}$, in order to generate two tones, one just above ($f_+ = f_0 + \delta f_{LO}/2$) and one just below ($f_- = f_0 - \delta f{LO}/2$) the detector resonant frequency. The modulation is carried out at about $1~kHz$, synchronously to the FPGA sampling of the signal. Thus, each raw data point, which is sent to the acquisition software at a rate of 23.842~Hz, is composed of the values $(I(t), Q(t))$ of each pixel, as well as the corresponding differential values
\begin{equation}
\left(\frac{dI}{df}(t), \frac{dQ}{df}(t)\right) = \left(\frac{I(f_+)-I(f_-)}{\delta f_{LO}}, \frac{Q(f_+)-Q(f_-)}{\delta f_{LO}}\right).
\label{eq:dIdQ}
\end{equation}
If a variation $(\Delta I(t), \Delta Q(t))$ is observed between successive points, it is possible to estimate the corresponding shift in the resonant frequency, $\Delta f_0(t)$, by projecting $(\Delta I(t), \Delta Q(t))$ along the gradient found using equation \ref{eq:dIdQ}:
\begin{equation}
\Delta \hat{f_{0}}(t) = \frac{\left(\Delta I(t), \Delta Q(t)\right)\cdot\left(dI/df(t), dQ/df(t)\right)}{\left(dI/df(t), dQ/df(t)\right)^2}\cdot\delta f_{LO}.
\label{eq:RFdIdQ}
\end{equation}
We use the name \emph{RFdIdQ} to refer to this estimate of $\Delta f_0$.

\begin{figure}[t!]
\begin{center}
\includegraphics[bb = 132 256 450 568,width=7cm, clip]{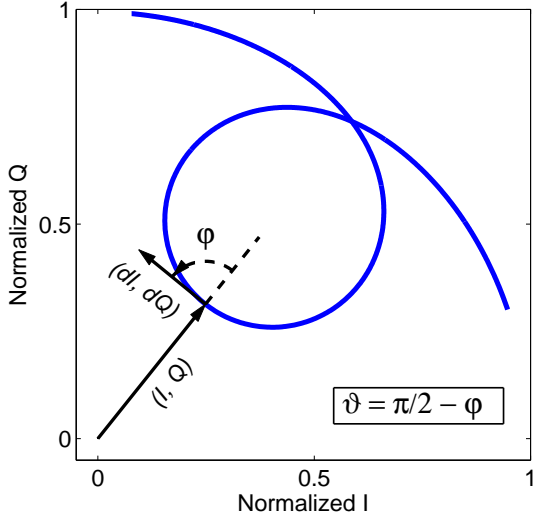}
\end{center}
  \caption{Representation in the I-Q plane of a sweep around a resonance. The measure of the angles $\varphi$ and $\vartheta$ can be carried out for each acquired point thanks to the modulated readout technique. On resonance, one gets $\vartheta = 0$}
\label{fig:angle}
\end{figure}

\subsection{Automated tuning procedure}
The modulated readout technique is the core of a fast and effective method of retuning the detectors that we have successfully implemented during the last two campaigns. When operating from ground, the variations in the background load due to the atmosphere can cause the resonant frequency of the detectors to vary by a substantial amount, introducing shifts that can be in some cases greater than the resonances themselves. This effect must be constantly monitored and, if needed, counterbalanced by changing the excitation tones, in order to always match the resonant frequency of each pixel, thus keeping the detectors near to their ideal working point.

The standard solution is that of performing full frequency sweeps before each on-sky observation, leading to a significant loss of observing time. The new tuning method is based on the measurement of the angle $\varphi$ between the vectors ($I$, $Q$) and ($dI/df_{LO}$, $dQ/df_{LO}$), as shown in figure \ref{fig:angle}. Thus, a single data point is now sufficient to retune the detectors without recurring to frequency sweeps. This leads to a crucial advantage in terms of observing time, especially in the case of medium and poor weather conditions. The new procedure reduces the required time for retuning by 75~\%.
Furthermore, in the case of altazimuthal maps, which are always composed of different subscans, the fast-tuning method makes it, in principle, possible to recenter the tones during the time spent by the telescope for changing the direction of its motion at the end of each subscan. Although we still have not made use of this approach, it does not pose any fundamental problem, and might prove highly effective especially in the case of large maps and long integration, in which case the sky conditions can change significantly between the start and the end of an observation.

The tuning process takes place as follows: from the angle $\varphi$ we calculate $\vartheta = \pi/2 - \varphi$, so that the new angle $\vartheta$ varies smoothly, decreasing across each resonance from $\pi$ to $-\pi$. After an initial frequency scan has been performed to find the resonances, we fix each excitation tone where $\vartheta=0$. At the same time, the slope of the curve $\vartheta(f)$, which is approximately linear around the resonance, is determined as  $\Delta\vartheta/\Delta f$. Once the tones are fixed, for each tone at frequency $f^i$ it is possible to continuously monitor the value of $\vartheta^i$. If the corresponding resonant frequency $f_0^i$ shifts, owing to changes in the optical load, this directly translates into a variation in $\vartheta^i$, from which it is possible to estimate the actual value of $f_0^i(t)$ as
\begin{figure}[t!]
\begin{center}
\includegraphics[bb = 98 305 490 509,width=8cm, clip]{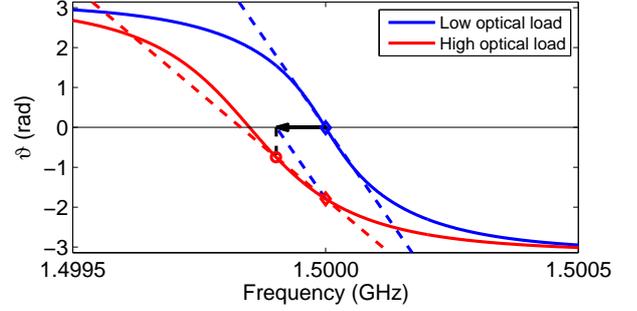}
\end{center}
  \caption{Representation of the tuning technique based on the measurement of $\vartheta$. A sweep (solid line) is carried out to look for the resonances where $\vartheta=0$. The tone is fixed at the corresponding frequency (diamond) and the slope evaluated (dashed line). If the optical load increases, $\vartheta$ changes allowing the appropriate correction $\Delta f^i$ (black arrow) to be evaluated. The new tone (circle) will be nearer to the actual position of the resonant frequency, $f_0^i$, and will be used to update the estimated slope of the $\vartheta(f)$ curve, thus making the iterative tunings increasingly more accurate.}
\label{fig:sweep}
\end{figure}
\begin{equation}
f_0^i(t)\simeq {f^i} - \frac{\vartheta^i(t)}{\Delta\vartheta/\Delta f}.
\label{eq:tuning}
\end{equation}
Although this relationship is not exact, mainly because the
changes in the optical load affect the $\vartheta(f)$ relationship and the
corresponding value of $\Delta\vartheta/\Delta f$, the results are very accurate,
provided that the shift in the resonance frequency does not exceed the
resonance width.

The results can be further improved by iterating the process. For this reason
every time that the telescope is not observing, we activate an automated tuning
procedure. This procedure repeatedly estimates the current value of $f_0^i(t)$
for each detector using equation \ref{eq:tuning}, adjusts the corresponding
tone accordingly (figure \ref{fig:sweep}), and updates the coefficient
$\Delta\vartheta/\Delta f$ by measuring the value of $\vartheta^i$ just before and
just after changing the frequency $f^i$. Thus taking the changes
in the slope of $\vartheta(f)$ into account, the excitation tone $f^i$ rapidly
converge to the correct resonant frequency $f_0^i(t)$. The procedure is then
halted as soon as a new observation starts.
In preparation for the NIKA open scientific pools, we will add an automatic check of the distance of each tone from its corresponding resonance. If this distance increases too much (something that would lead to incorrect data), the tone is automatically recentered on the resonance. The process is thus not continuous but only carried out in case of need. Under good sky weather conditions, even very long scans can be carried out without having to change the tones. This automatic tuning process will allow astronomers to observe throughout the future campaigns with no interventions of the supporting NIKA team.

%% file: 04_focal.tex
The frequency multiplexing of \NIKA\ prevents us from knowing {\it a priori} the pointing
direction of each detector on the sky. This has to be determined on
astronomical observations. We therefore scan a strong astromical source with the
entire focal plane. The calibration source, typically a planet (Mars, Uranus,
Neptune), is small compared to our beam and can be considered as a point
source. Indeed, during the 2012 observing campaign, the angular
diameter of Uranus was 3.54~arcsec. The convolution of the corresponding disk with a Gaussian beam of
13.5 and 18.4~arcsec FWHM would broaden our beam by only 0.17~arcsec at 1.25~mm and 0.12 at 2.14~mm.

The source is raster-scanned at 35~arcsec/s, and each subscan is 420~arcsec
long, centered on the source as the latter is being tracked by 
the telescope. This scanning speed provides about 10 and 12 points per FWHM at
  1.25~mm and 2.14~mm resp., hence providing excellent beam sampling. To
have a clean determination of the beam parameters (position, width, and
orientation) of each KID, we proceed in two steps.
 
\begin{enumerate} 
\item We apply a median filter of approximately 5~FWHM of width to the detector timelines to
  subtract most of the atmospheric signal and the low-frequency, correlated
  electronic noise while preserving most of the planet signal (less than 1~\% lost at scales 
  smaller than 2.5 x FWHM). These timelines
  are then projected onto individual maps per detector, and a Gaussian ellipse is
  fitted on the source. Centroid position provides a first estimate of the pointing
  parameters of each pixel. The amplitude of the Gaussian gives a calibration in
  Jy/Hz, where Hz represents the shift in resonant frequency 
  for each detector. If the atmospheric absorption is known and accounted for, this
  becomes a determination of the absolute calibration of each detector. If not,
  this can still be considered as a cross-calibration of KIDs all at once, and
  they can be combined accordingly into a single map of the source. This map
  provides the position of the source in sky coordinates.
\item With this information in hand, we can flag out the source in all KID
  timelines and build a template of the low frequency part of the signal (mostly
  sky noise and electronics noise) using, at each time, all detectors that are
  far from the source (typically further than a few beam FWHM,
  e.g.~30~arcsec). This template is then subtracted from each KID timeline. This
  leads to a clean determination of the planet signal with no filtering. Timelines
  are then projected a second time onto individual maps per KID, and a Gaussian
  elliptical fit provides the final pointing parameters and FWHM estimates.
\end{enumerate}

\begin{figure}[t!]
\includegraphics[width=4cm]{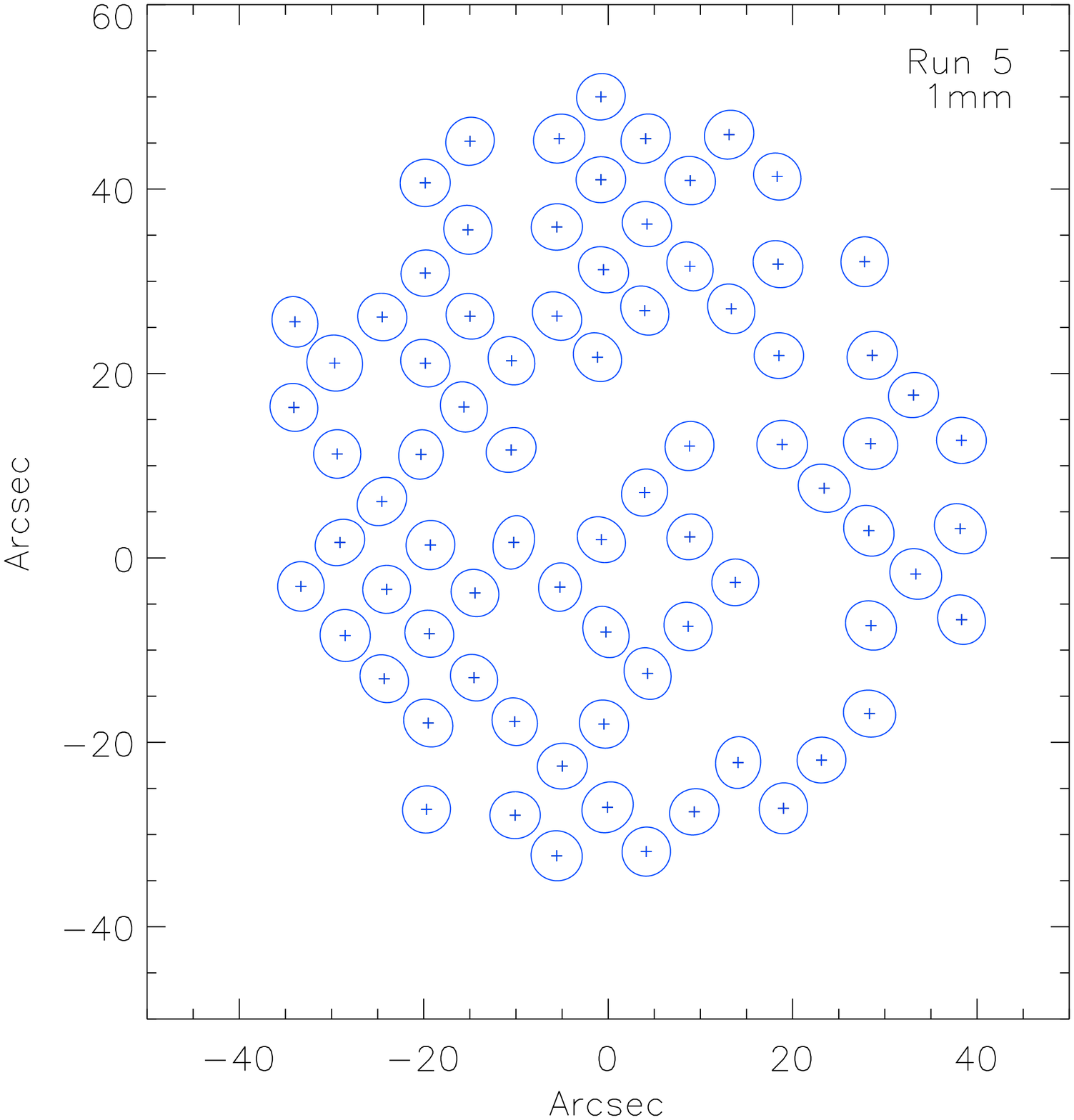}
\includegraphics[width=4.2cm]{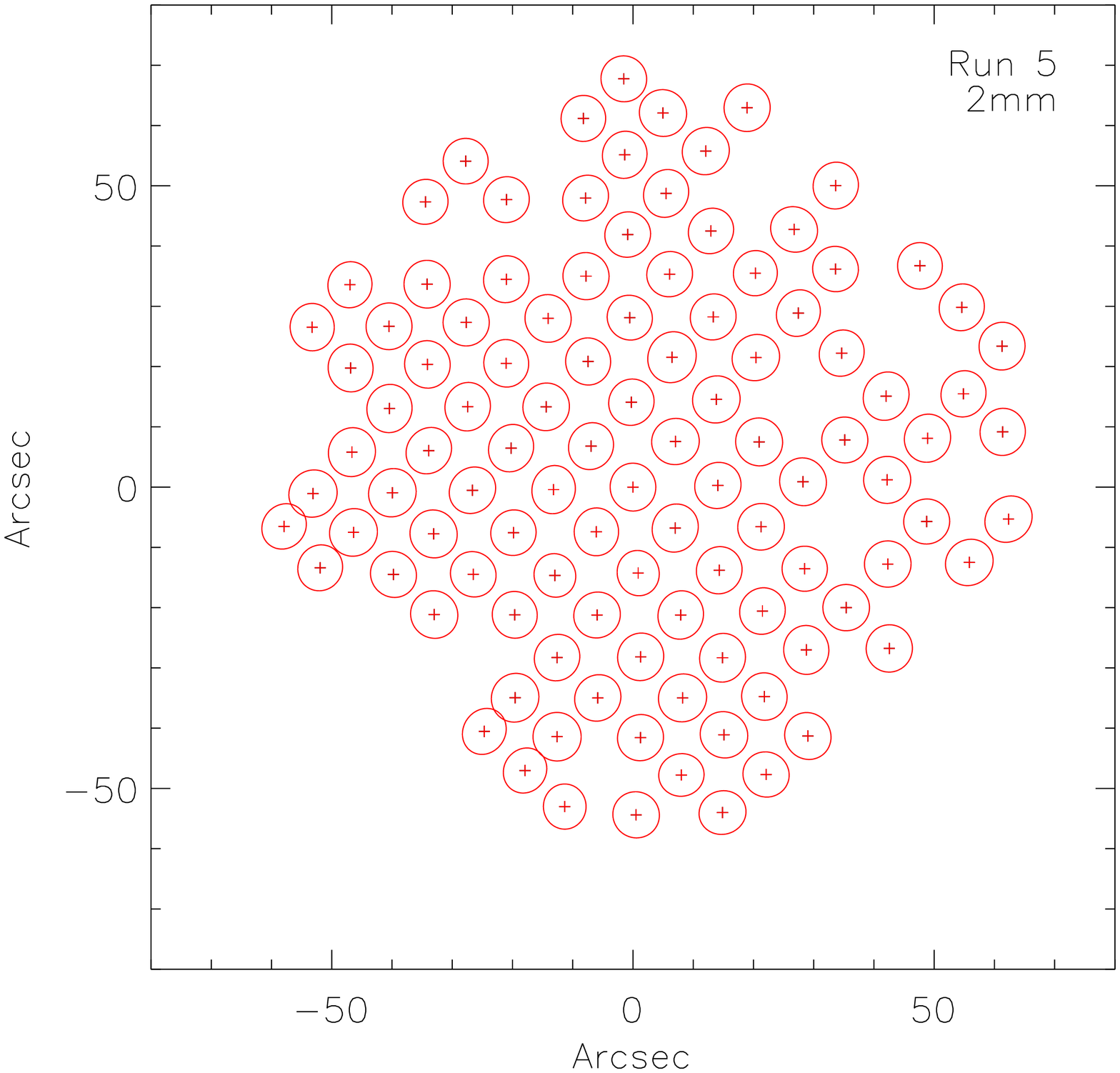}\\
\includegraphics[width=4.3cm]{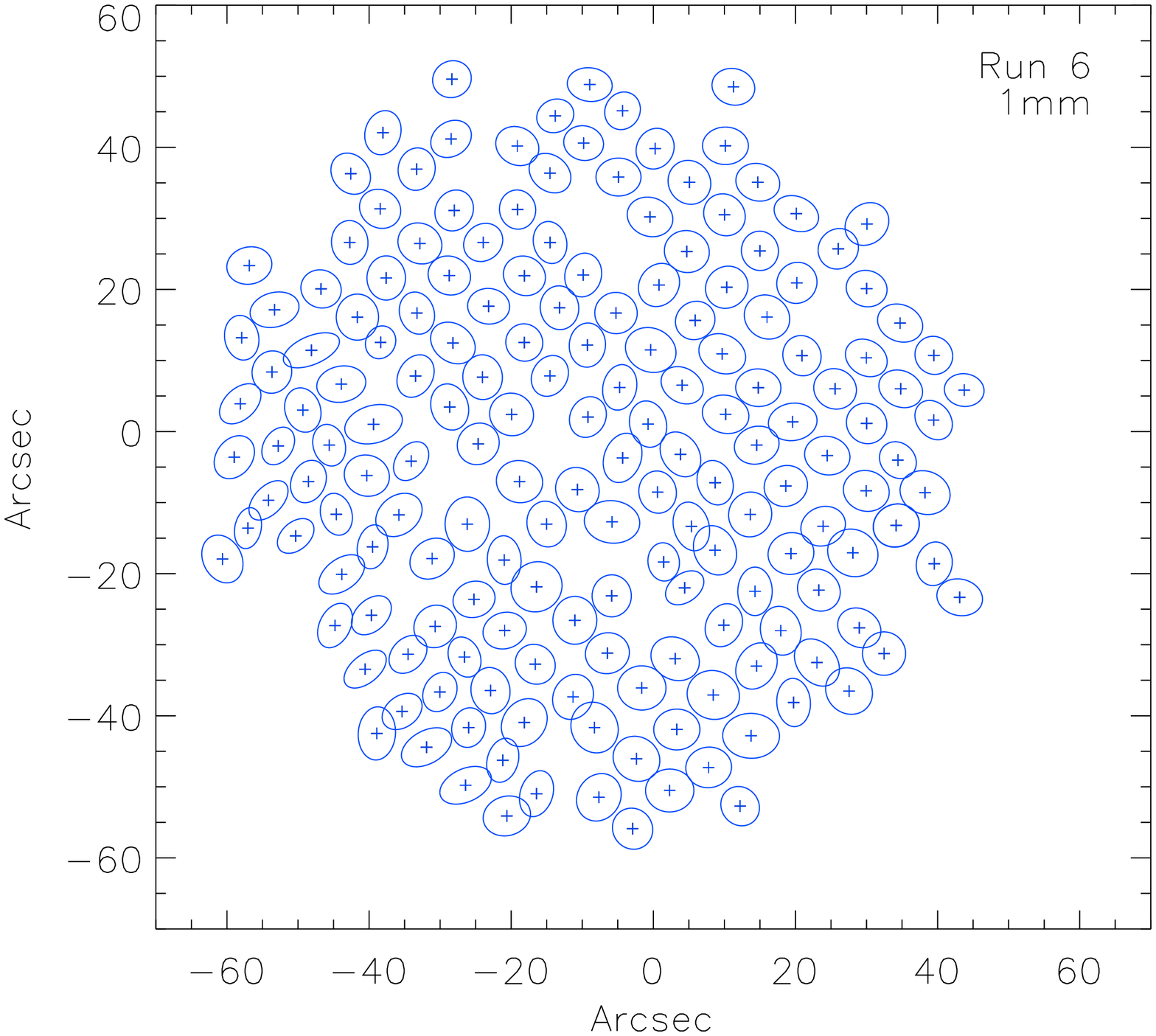}
\includegraphics[width=4.1cm]{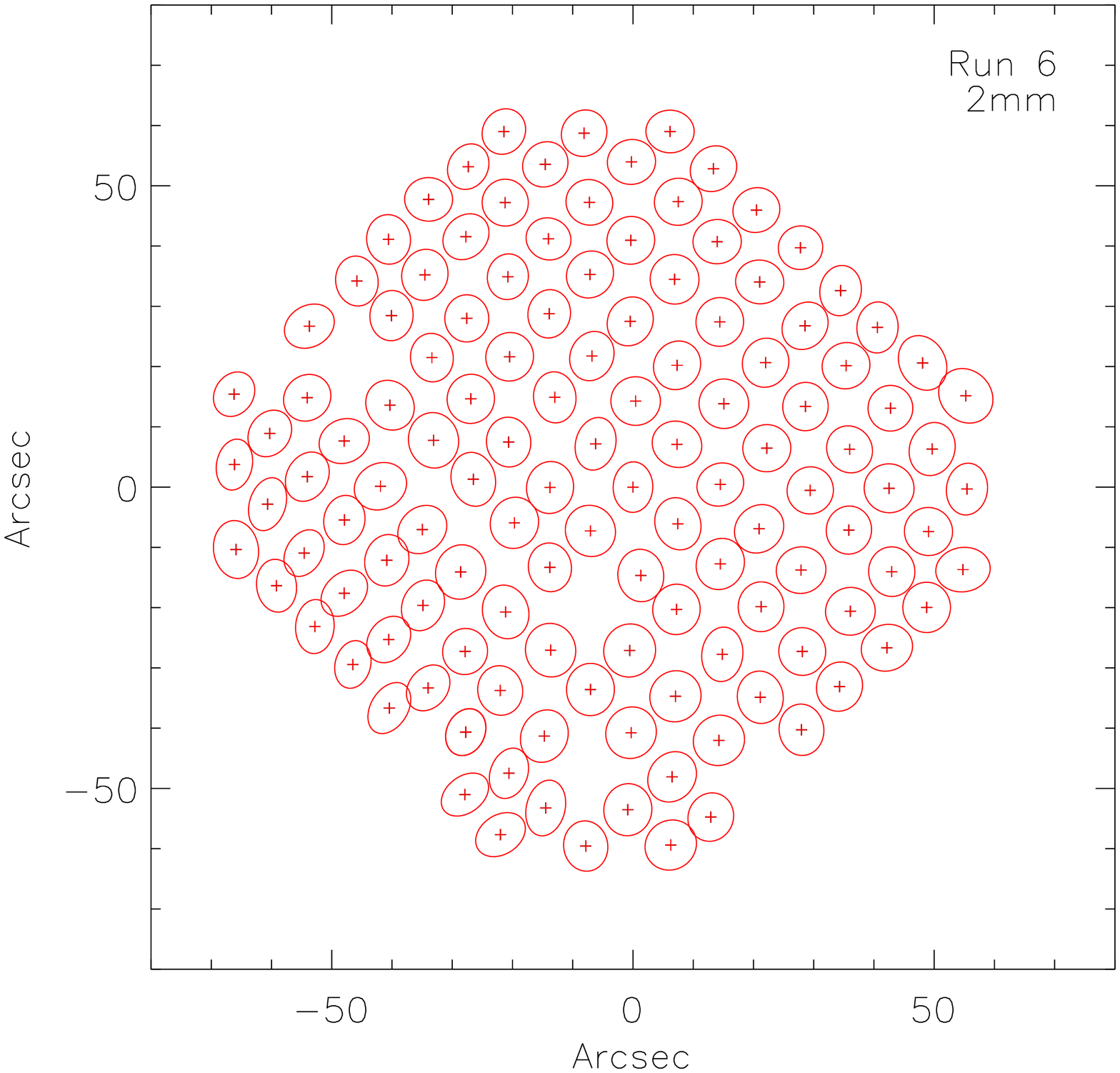}
\caption{Detector sampling of the FOV at 1.25 mm (left) and 2.14 mm (right) for the 2012 (top) and 2013 (bottom) observing campaigns. Beam
  pattern contours have a diameter of $1\sigma=FWHM/\sqrt{8\ln 2}$.}
\label{fig:run6_fpg}
\end{figure}

\begin{figure}[t!]
\includegraphics[width=4.2cm]{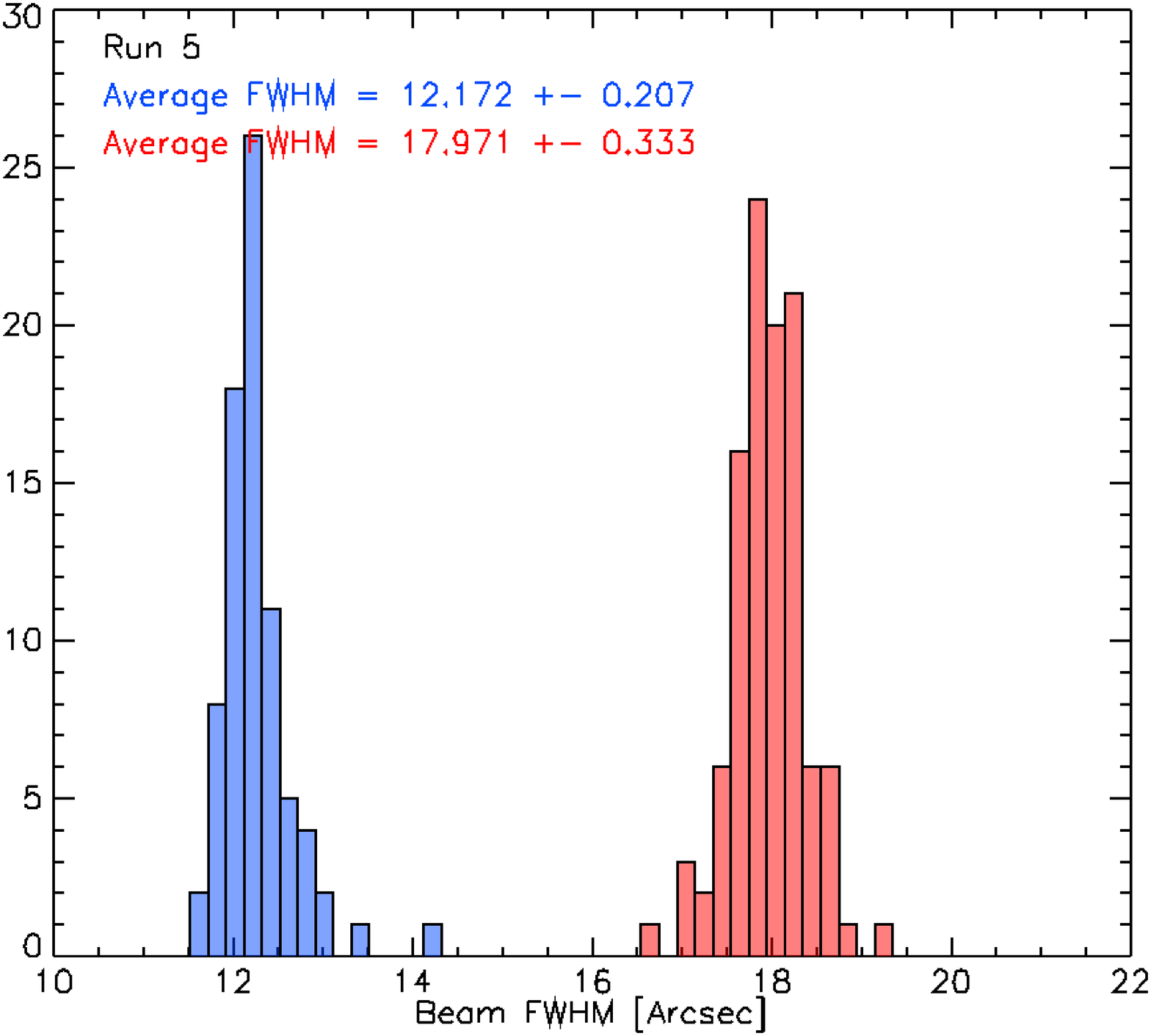}
\includegraphics[width=4.3cm]{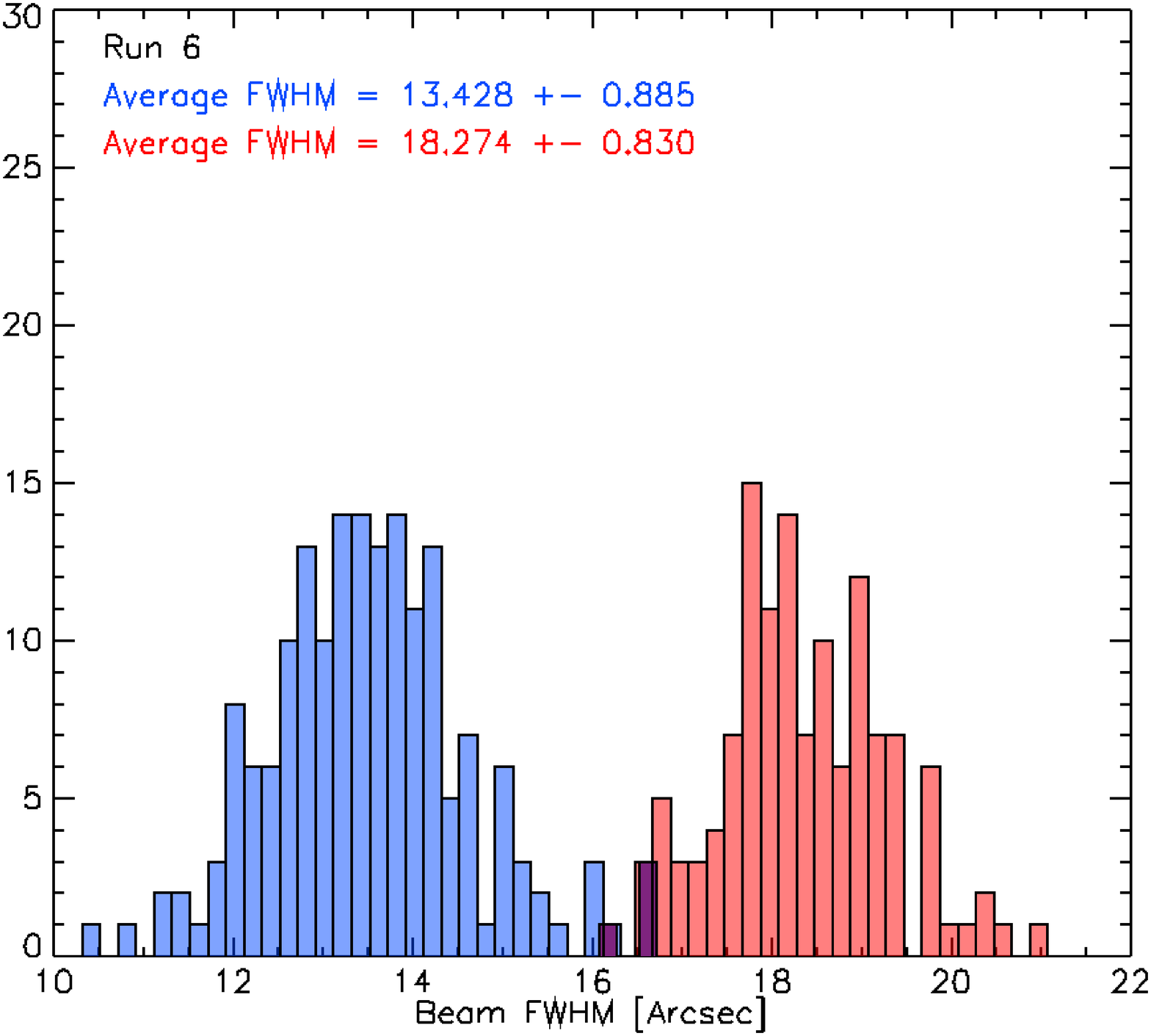}
\caption{Beam FWHM distribution at 1.25 (blue) and 2.14 (red)~mm for the 2012 (left) and 2013 (right) observing campaigns. 
The distribution of FWHMs was of lower quality during the 2013 observing campaign owing to
  poor weather conditions.}
\label{fig:run6_beam_stats}
\end{figure}

For illustration, Figures~\ref{fig:run6_fpg} and \ref{fig:run6_beam_stats} show
maps of the 1.25~mm and 2.14~mm detector images and the distribution of the
determined beam widths for the 2012 and 2013 observing campaigns. Several scans on
calibration sources are performed during the campaign to improve on the
reconstruction of the pointing. For the 2012 observing campaign, with typically
six such scans, we observed a median variation of the detector pointing
directions of 3.4~arcsec at 1.25~mm and 3.2 at 2.14~mm.
Although beams were observed to be
a bit larger and asymmetric during 2013 observing campaign because of the weather conditions\footnote{Under turbulent sky, the anomalous refraction is larger and spreads the incoming light, hence enlarging the image.}, the number of valid pixels increased and the FOV was almost fully sampled. For the
2012 observing campaign, the average FWHM were $12.3\pm 0.2$~arcsec at 1.25~mm and $18.1 \pm 0.3$~arcsec
at 2.14~mm. For the 2013 observing campaign, they were $13.5\pm 0.9$ at 1.25~mm and
$18.4\pm0.8$~arcsec at 2.14~mm.
The combined beam is larger than individual detector beams because of a possible small offset in the reconstructed detector position in the FOV. 

%% file: 05_toiproc.tex

As discussed in section \ref{REU}, the NIKA data are acquired at 23.842~Hz,
and each pixel is fed a single tone. For each sample and for each KID, the in-phase (I), the
quadrature (Q), and their derivatives (dI and dQ) of the transfer function
of the feed line and the pixel are recovered.
The variation in resonance frequency for each KID in the array is obtained by applying
the optimization procedure, \emph{RFdIdQ}, discussed in Sect.~\ref{sec:rfdidq}.
We developed a dedicated
reduction pipeline to calibrate, filter, and process data onto sky maps.  
The main steps of the processing are:
\begin{itemize}

\item {\bfseries Read raw} : raw data and the main instrumental parameters
including FOV reconstruction and atmospheric
opacity for each observational scan. The data are ordered per kid and
regularly sampled with time. This is called TOI (time ordered information).

\item {\bfseries Flags}: Bad detectors are flagged based on the
  frequency range not being large enough to have a low cross-talk level. 
  Noisy or badly systematic, affected detectors are flagged depending on the statistical 
  properties of their noise (such as Gaussianity, stationarity, noise jumps). 
  Saturated and off-tone KIDs are also flagged out.

 \item {\bfseries Filtering}: Frequency lines produced
  by the vibration of the cryostat pulse tube are flagged and removed with
  dedicated Fourier space filtering.

\item {\bfseries Cosmic rays}: We detect glitches in the raw data. Since
  1.25~mm and 2.14~mm KID arrays are different (in terms of size of the
  pixels and thickness of the substrate), the rate of observed glitches varies
  between them. We observe about six glitches/min in the 2.14~mm array
  and about four for the 1.25~mm array. The observed glitch rate is in
  good agreement with the expected cosmic ray flux at the IRAM telescope,
  which is essentially composed by muons with a rate of about 2 $\mathrm{events / cm^2 /
    min}$ \cite{Ramesh2012}. The time response of the KIDs is hundreds of
  $\mathrm{\mu s}$, which is negligible compared to the NIKA acquisition rate,
  therefore each cosmic ray hit only affects only a single sample. Glitches
  are removed from the TOI by flagging peaks at 5\,$\sigma$ level. Then the
  TOI is linearly interpolated in order not to perturb the decorrelation
  method. Flagged samples are not projected onto the sky.

\item {\bfseries Calibration of the TOI}: The absolute calibration is applied
  to these TOIs and an atmospheric absorption correction performed (see section
  \ref{skydips}).

\item {\bfseries Atmospheric and electronic noise decorrelation}: Depending
  on the scientific target, two basic decorrelation methods have been
  tested:

  1) dual-band decorrelation: spectral decorrelation is performed
  to recover the diffuse thermal SZ effect (see \cite{2013arXiv1310.6237A} for more detail) 
  in clusters of galaxies. We capitalize on the significant difference between the thermal SZ associated signal and
  the two frequency channels. In particular, at 240 GHz the thermal
  SZ emission is negligible to first order, and we can use this channel to obtain a template of the
  atmospheric emission.

  2) single-band decorrelation: a sky noise TOI template is produced by
  averaging all TOIs of a single array. This template is scaled for each
  detector and subtracted from individual TOIs. Variations on this method have
  been tested. They involve masking the point source if bright enough, or
  masking an extended source if detected in a previous iteration of the
  map-making process or filtering the sky noise template.
  In particular, we consider an iterative procedure. A first sky map
  is constructed as discussed above. From this map we construct a mask
  of the source in the TOI. We construct a new TOI sky noise template, avoiding
  samples within the mask. Only a few iterations are needed to converge.

So far, the electronic noise has not been dealt with. In general, 
the cross-talk can be due either to resonator coupling or to electronics cross-talk. 
The residual estimated cross-talk in the data is about 2~\%. Decorrelation process 
reduces, or in some case completely removes, the cross-talk. This can limit the sensitivity of
the final sky maps. We are currently investigating how to correct for this by relying on the use of measurements at frequencies that are off-resonance.


\item {\bfseries map making} : We project and average the TOIs from
all KIDs of a same frequency band on a pixelized map. We do not project flagged data,
and each detector sample is weighted by the inverse variance of the detector
timeline. The latter is estimated far from the source. This weighting scheme
does not average down any residual correlated noise coming from the sky, for
instance. The error bar on flux measurement that we quote is therefore not
computed analytically from these weights but rather from the noise on the map directly.

 Uncertainties on final maps are also computed.

\end{itemize}


%% file: 06_photcal.tex
The variety of \NIKA\ scientific targets going from thermal SZ observations to
dust polarization properties requires an accurate calibration process able
to define the impact of systematic errors on the final sky
maps as well as possible. The overall calibration uncertainty for point sources on the final data 
at the map level is estimated to be 
15~\% for the 1.25~mm channel and 10~\% for 2.14~mm channel.
In the following, a list of the principal error sources, which have a
direct impact on the total calibration uncertainties, is presented. In Table
\ref{tab:table_err}, we quantify these contributions.

\begin{figure*}[t!]
\begin{center}
\includegraphics[width=7cm]{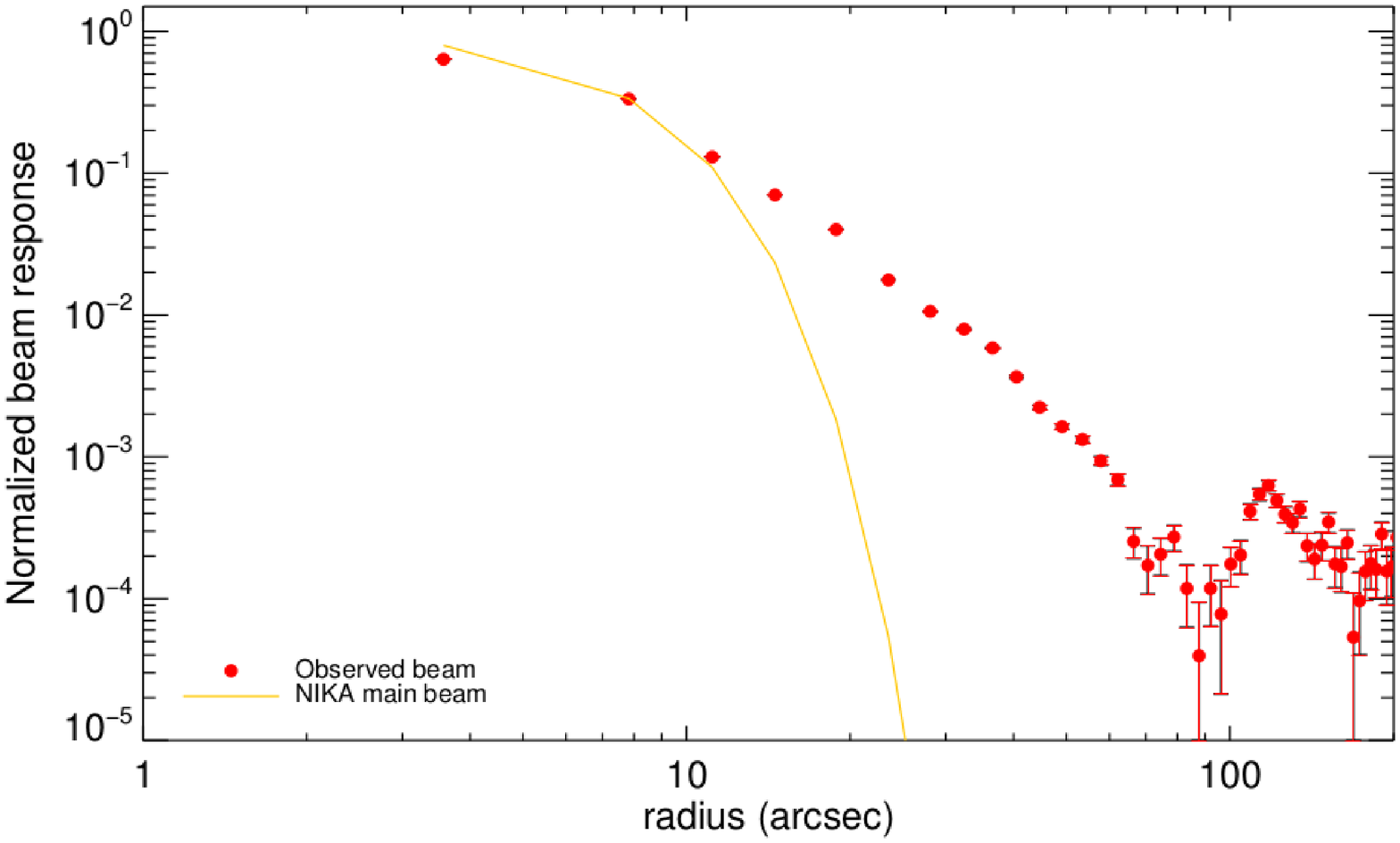} 
\includegraphics[width=7cm]{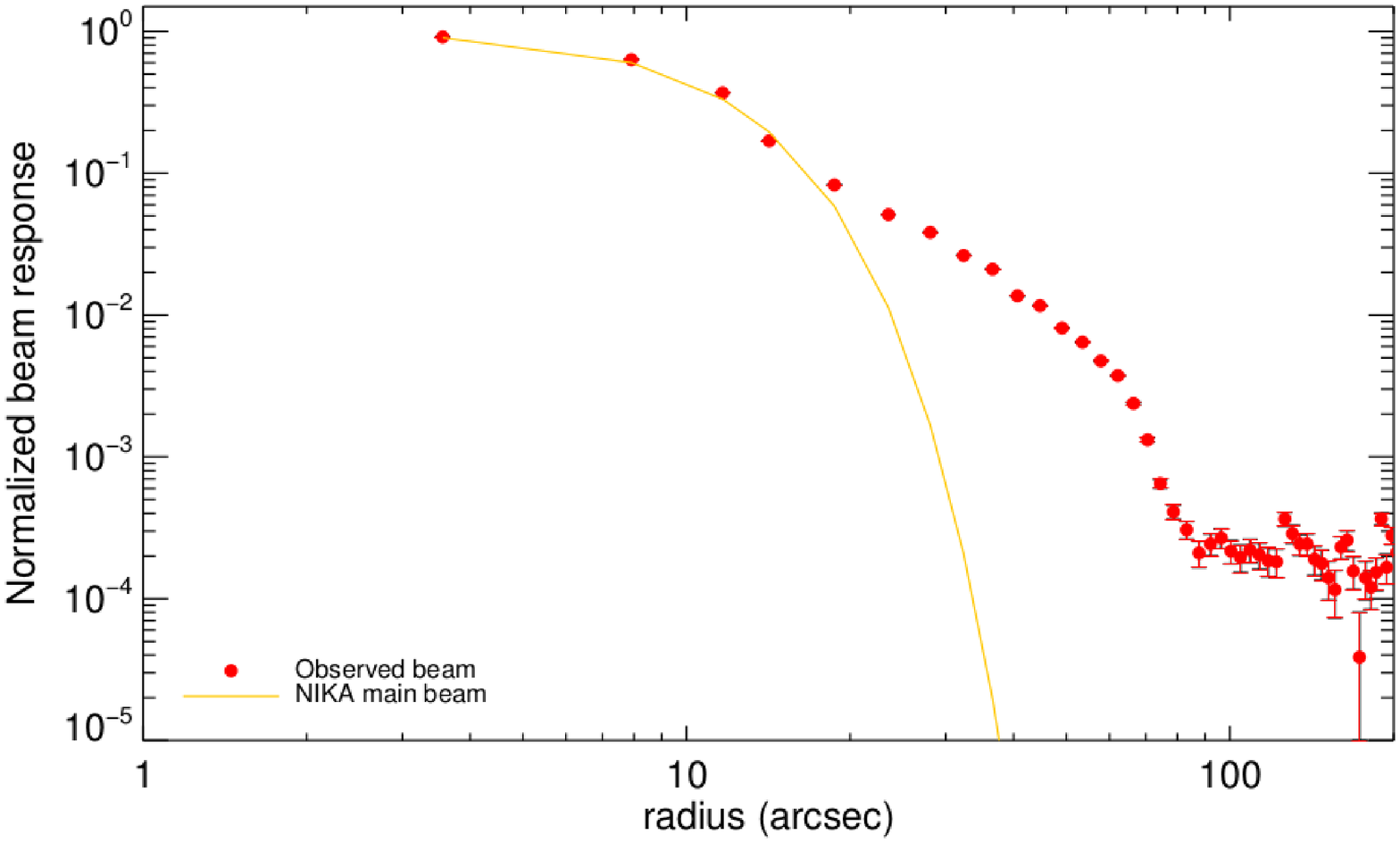} \\
\includegraphics[width=7cm,trim=0cm 0.35cm 0cm 0cm,clip]{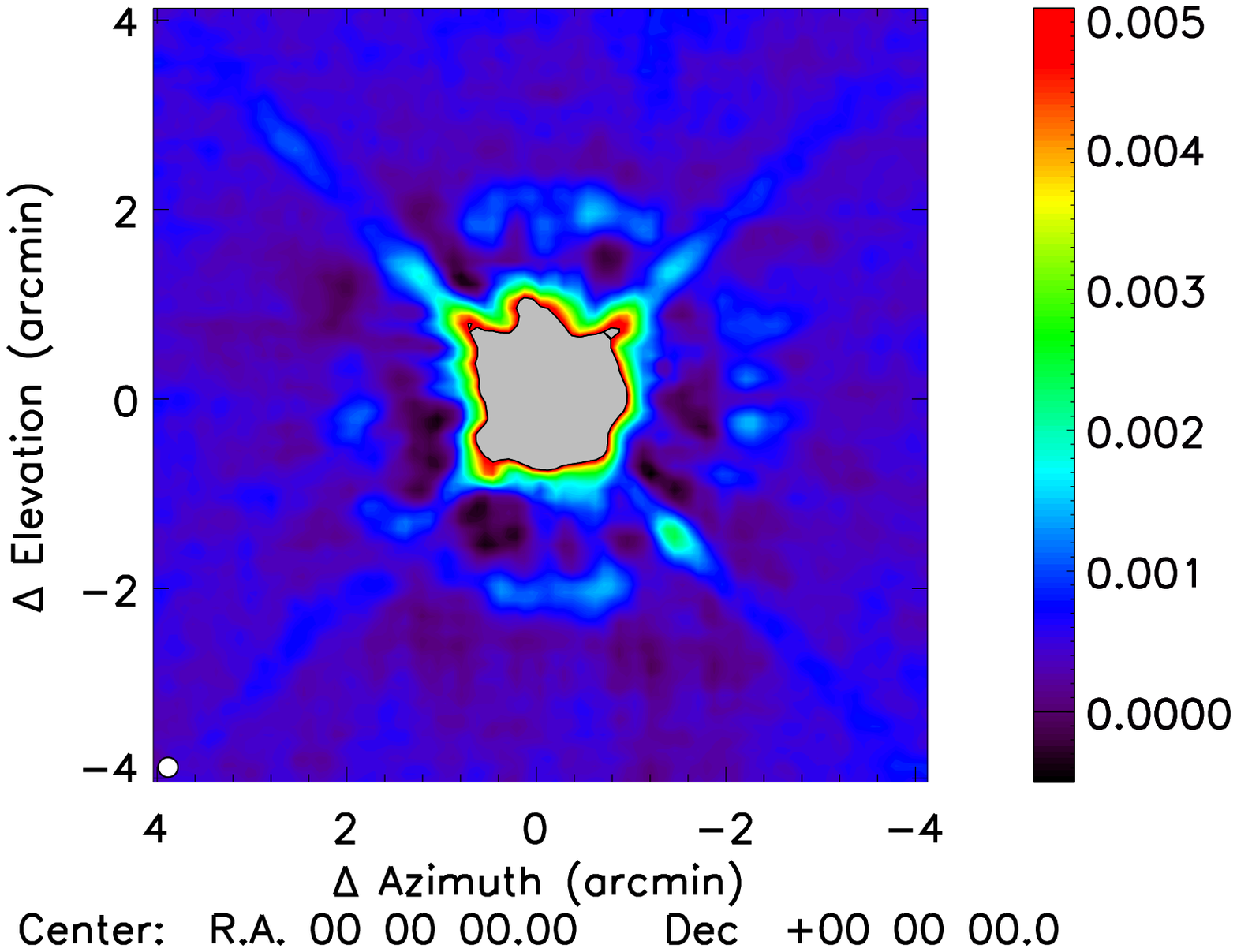} 
\includegraphics[width=7cm,trim=0cm 0.35cm 0cm 0cm,clip]{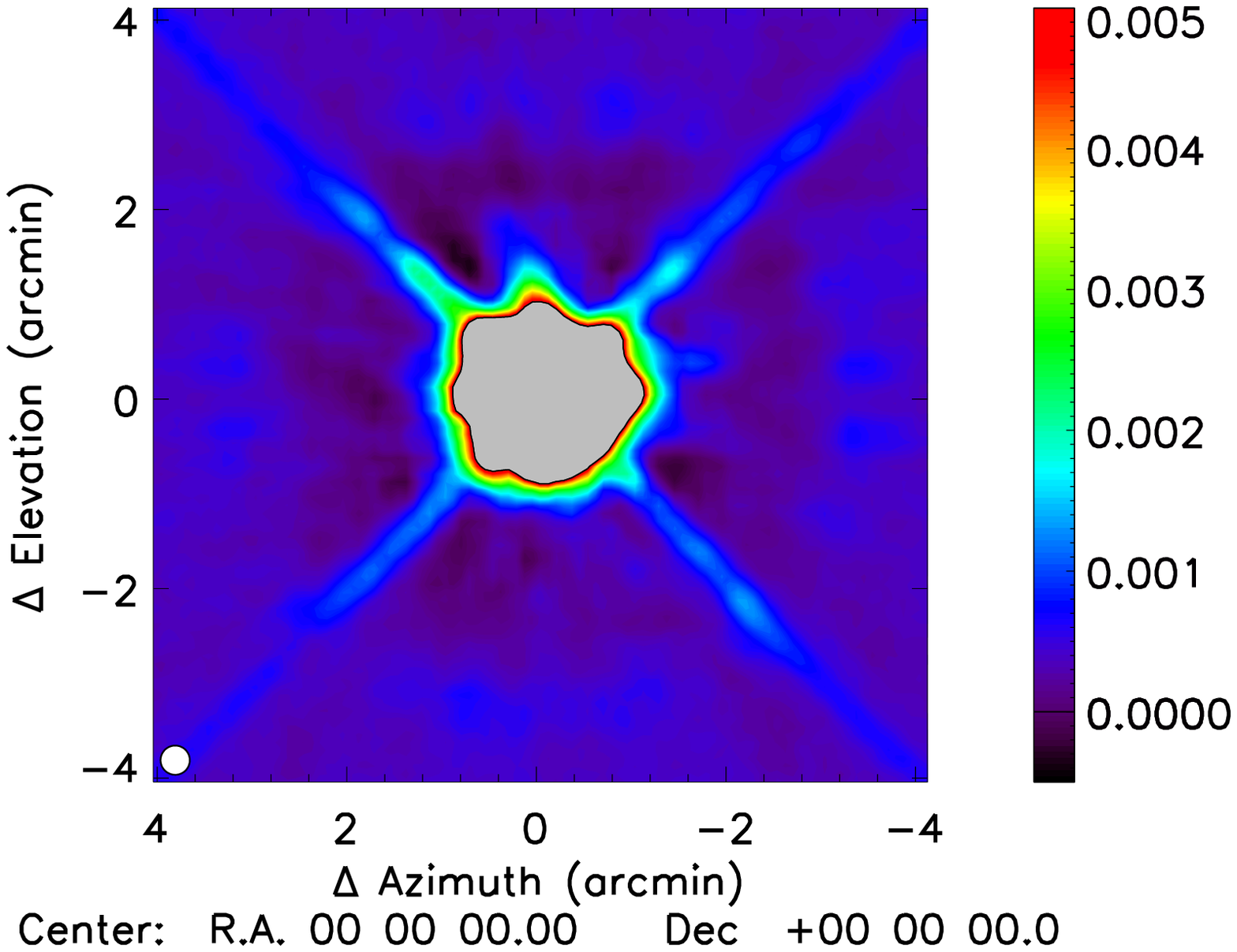} 
\end{center}
\caption{Top panel: Radial profile of the \NIKA\ beam pattern (red dots) including the secondary beam contribution for the 1.25~mm channel (left) and 2.14~mm channel (right) in the case of Uranus observations. We are able to measure the beam pattern profile up to scales of about 100 arcsec. At larger distances we do not have enough signal-to-noise. For illustration we also show the \NIKA\ measured primary beam. 
 Bottom panel: Map of the far side lobes for 1.25~mm channel (left) and for 2.14~mm channel (right). The maps are derived using observation of Saturn. These maps are compatible with the  the 2D structure of the beam pattern of the IRAM 30~m telescope described in \citep{Greve,kramer}). The diffraction ring seen at 1.25~mm at about 2~arcmin radius, corresponds to the diffraction ring  due to panel buckling. The spider supporting the secondary mirror of the telescope is visible to a level of about -30~dB.}
\label{fig:fsl}
   \end{figure*}

\begin{itemize}

\item {\bfseries Primary calibrator}:  Uranus was selected as the primary calibrator 
for the \NIKA\ absolute calibration. The absolute calibration factor is derived by fitting 
a Gaussian of fixed angular size on the reconstructed maps of Uranus. 
The flux of Uranus is deduced from \cite{moreno2010,planckii}  using a frequency-dependent
model of the planet brightness temperature and integrating over the \NIKA\ bandpasses. We obtained brightness temperatures of 113~K at 2.14~mm and 94~K at 1.25~mm. This model is accurate at the 
level of 5~\%. 

\item {\bfseries Elevation dependent gain}: The IRAM 30~m telescope has 
a gain that changes with the elevation. This is because the antenna was not designed in a complete homology way. 
According to the design and measurements \citep{Greve, kramer}, the peak
of the elevation gain curve is obtained for an elevation of 43~degrees. 
Elevation gain variations have been corrected for in the data analysis.

\item {\bfseries Spectral response}: The bandpasses as described in section
  \ref{instr} are obtained as an average of all pixels for each channel. This
  leads to a contribution to the total calibration error due to the dispersion
  of the measured bandpasses and to the slightly different response of the
  pixels.

\item {\bfseries FOV reconstruction}: To recover the
  pointing direction of each pixel, a focal plane reconstruction via planet
  scans is used. The accuracy of this technique has been described in Section~\ref{focal}.

\item {\bfseries Secondary beam fraction}: Planet observations were used to
  measure individual pixel primary beams as well as for the FOV
  reconstruction. The beam width is obtained by fitting a Gaussian on the
  planet maps. We have observed that variations in the atmospheric conditions
  lead to changes on the beam width. Since we calibrate assuming a fix beamwidth,
  these variations induce a systematic error in the final calibration.

\item {\bfseries Opacity correction}: The sky maps are
  corrected for the atmospheric contribution rescaling the observed signal by
  what that would be obtained in the absence of atmosphere. This
  is achieved via the elevation scan technique (skydip). The \NIKA\ skydip
  procedure was successfully tested during the last two \NIKA\ observing
  campaigns, and it produced a low-level dispersion of the derived opacity at
  different elevations.

\item {\bfseries Data reduction filtering}: 
The main steps in the data processing 
have been discussed in section \ref{dataprocessing}. The estimated systematic error 
introduced by the data reduction filtering (in particular due to the common modes subtraction) 
is estimated at 5~\% for both channels. This was computed from the dispersion between different
processing modes.

\end{itemize}

In the following we discuss the secondary beam fraction
contribution and the atmospheric absorption correction in more detail.

\subsection{Secondary beam fraction}
The secondary beam fraction needs to be estimated and accounted for in the case of extended sources. 
This is known by comparing the \NIKA\ beam pattern with the beam pattern of the 30~m telescope as measured by \cite{kramer} (K13) on the Lunar edge using the EMIR receiver. For practical purpose we have divided the beam into three regions: short angular scales correspond to the main beam, intermediate angular scales corresponding to the first error beam, and large angular scales assimilated to far side lobes of the 30m telescope. To compute the main beam we performed a Gaussian fit to the full observed NIKA beam pattern. The best Gaussian fit is shown in yellow. The \NIKA\ main beam is consistent with the one of K13 because we expect the main beam to be defined by the diameter of the 30~m telescope. In the same way far side lobes, which are expected to come mainly from the second and third error beams of the 30~m telescope caused by small adjustment errors of the panels and their frames, are also consistent between \NIKA\ and K13. However, the first error beam observed in the \NIKA\ beam pattern is larger than the one modeled by K13. This is not unexpected since the observations were carried out at a different time of day and under different weather conditions.
In Fig \ref{fig:fsl}, we show profiles and maps of the beam emphasizing the contribution of the secondary beam for both \NIKA\ channels. The results on the estimation of its contribution are presented in Table~\ref{tab:table_err}.

\subsection{Atmospheric absorption correction using a \Skydip\
  calibration}\label{skydips}

\begin{figure}
\begin{center}
\includegraphics[width=8cm]{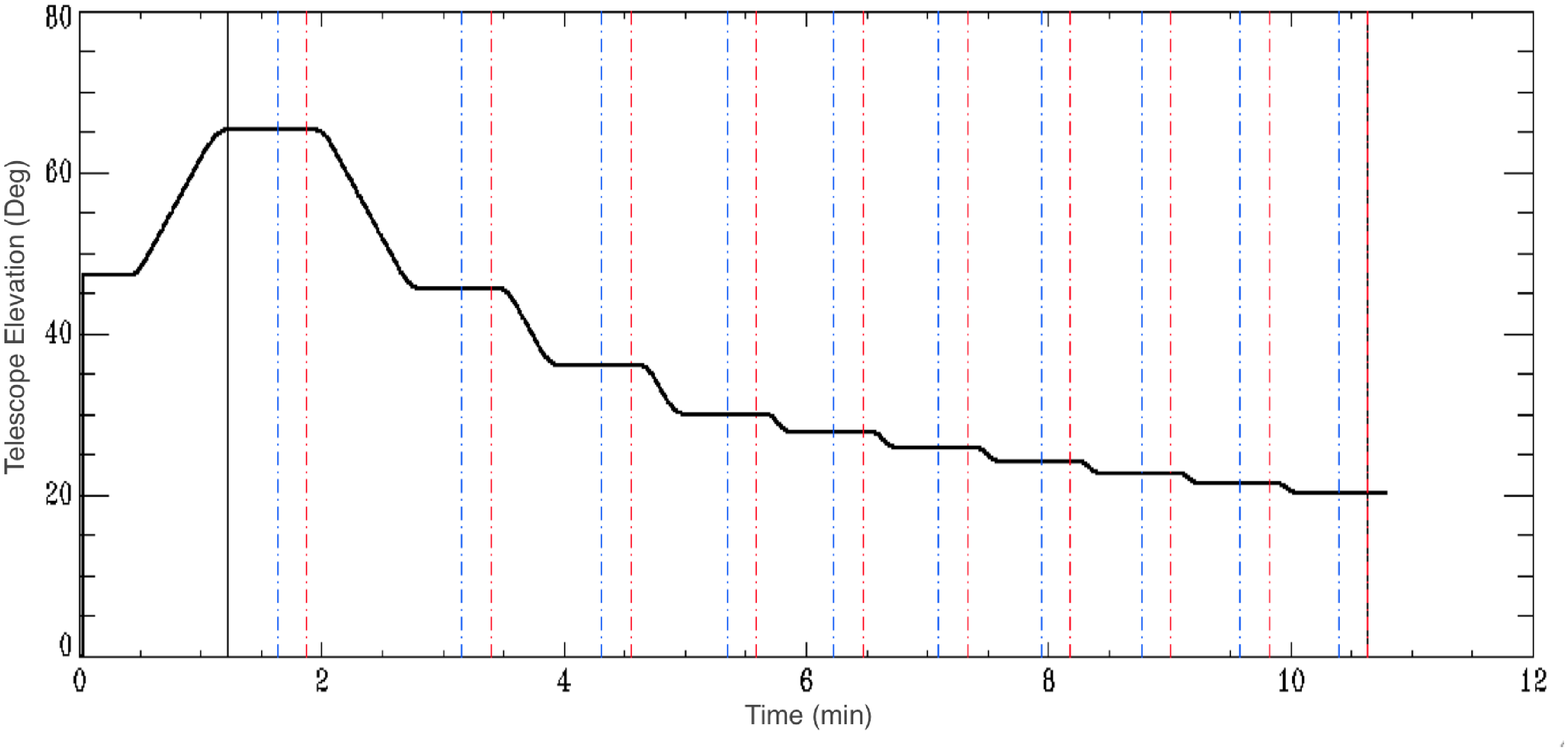} 
\includegraphics[width=9cm]{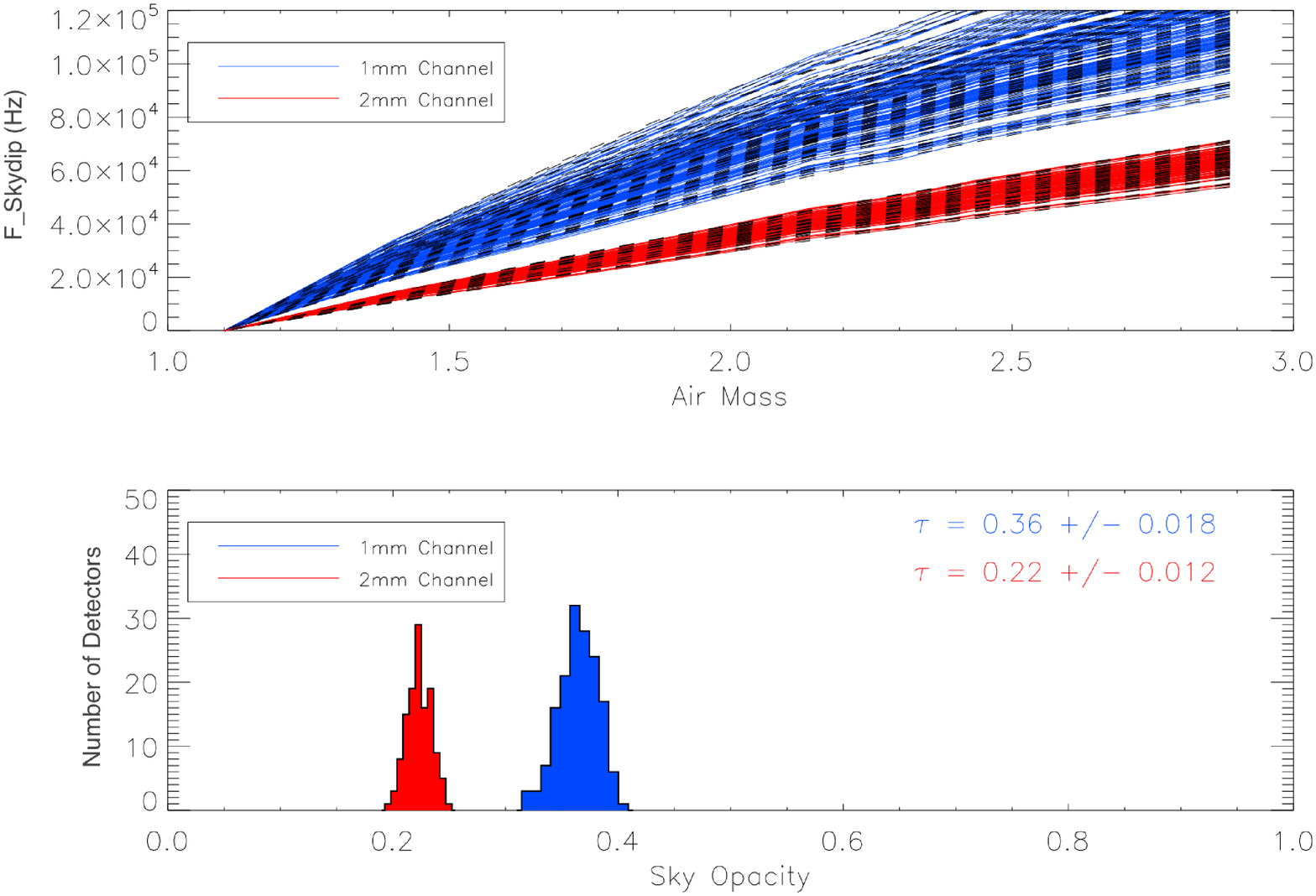}
\end{center}
\caption{Top: telescope positions during an elevation scan procedure: 10 steps in elevation have been performed without changing the azimuthal position. Data for absolute calibration are taken in the region
  between the blue and the red lines. Middle plot: signal in Hz of each valid
  KID (blue for 1.25~mm detectors, red for 2.14~mm) as a function of air mass, together with the fitted model (black dotted lines). Bottom plot: Histogram of the deduced
  integrated in-band opacities for the 1.25~mm and 2.14~mm channels.}
\label{fig:skydip}
\end{figure} 


In the previous observing campaigns, the atmospheric absorption correction
was made using the IRAM tau-meter that performs elevation scans continuously
at a fixed azimuth at 225~GHz. To derive the opacity at the exact position of
the scan and at the same frequencies as \NIKA\, we implemented a
procedure that uses the \NIKA\ instrument itself as a tau-meter. For the
last 2012 and 2013 observing campaigns, the \NIKA\ atmospheric calibration
consisted in measuring the variation in the resonance frequencies of the
detectors versus the airmass via elevation scans (\emph{skydip} \cite{dicke})
from 65 to 20 degrees above the horizon. The procedure is based 
on the idea that the KIDs response (the change in resonant frequency for a given 
change in absorbed power) is a constant property of each detector. This has been
demonstrated in laboratory under realistic conditions (with an optical load changing 
between about 50~K and 300~K, \cite{monfardiniLTD})

\begin{figure}[b!]
\begin{center}
\includegraphics[width=8cm]{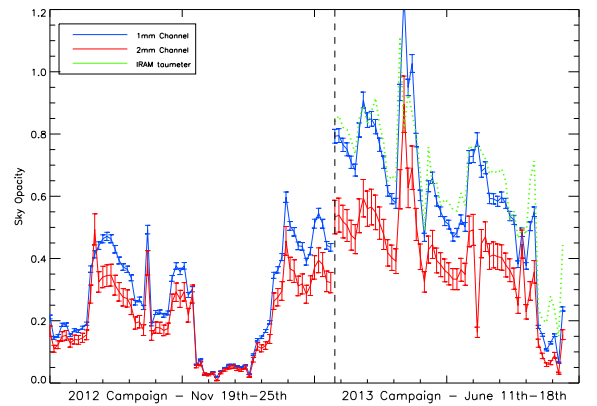}
\end{center}
\caption{Atmospheric opacity evolution for the \NIKA\ 2012 and 2013 observing campaigns 
calculated from the \Skydip\ analysis. The error bars were estimated by
analysing different \Skydip\ observations. During the 2013 observing campaign, the agreement with 225~GHz IRAM taumeter is good.}
\label{fig:op}
\end{figure}


During \Skydip\, the telescope performs ten elevation steps each corresponding to variations of 0.18 in air
mass. We perform a tuning of the readout electronics before acquiring a useful signal. For
each step we have 22 seconds of useful signal at a given elevation. This is shown
in the top plot of  Figure~\ref{fig:skydip} where we present the observed elevation as a function
of time during the \Skydip\ procedure.

We expect the acquired useful signal per detector to respond to the airmass as:

\begin{equation}\label{eq:skydip}
F^{Ground}_{skydip} = F_0 + C T_{atm}[1 - e^{- x \tau_{skydip}}].
\end{equation}

Here, $F^{Ground}_{skydip}$ is the acquired signal corresponding to the
absolute value of the shift in the frequency tone for each pixel, $F_0$ is the
instrumental offset corresponding to the frequency tone excitation for the
considered pixel for zero opacity, $C$ the calibration conversion factor in
$\mathrm{Hz/K}$, $T_{atm}$ (in Kelvin) is the equivalent temperature of the
atmosphere, $\tau_{skydip}$ the sky opacity during the skydip (at the
wavelength of the fit, either 1.25 or 2.14~mm), $x$ the airmass 
with $x = sec(\delta)$  where $\delta$ is the average elevation
of the telescope during the scan. By performing a fit of the three
parameters, $F_0$, $C$, and $\tau$, for all valid detectors at the same
wavelengths, we obtain a common sky opacity at zenith during the skydip. We
can rerun the fit on each detector, assuming the common $\tau$ value, and get
the coefficients $F_0$ and $C$ per detector. In the middle and bottom panels of Fig.~\ref{fig:skydip} we
present the main results of the data analysis of one \Skydip\ performed during the
2013 campaign.

\begin{table*}
\begin{center}
\begin{tabular}{ccc}
\hline
\hline
Systematics & 1.25~mm Channel error &  2.14~mm Channel error  \\
\hline \hline
Secondary beams fraction$^*$  (cuts at $30^{"}$, $60{"}$, $90^{"}$) & 25\%, 41\%, 43\%   & 9\%, 30\%, 33\% \\
Primary calibrator &  5\% & 5\% \\
Elevation dependant gain correction$^{**}$ &  20\% & 10\% \\
Spectral response &  2\% & 1\% \\
FOV reconstruction & 3.4 arcsec & 3.2 arcsec \\
Opacity correction &  5\% & 6.5\%  \\
Data reduction filtering (on point sources)  &  5\% & 5\%  \\
\hline
\bfseries{Overall calibration}  & \bfseries{15\%} & \bfseries{10\%} \\
\hline \hline
\end{tabular}
\end{center}
\caption{Different contributions to the total calibration error of the \NIKA\ data. $^*$  is estimated by measuring the main beam volume ($2\pi \sigma^2$) over the integral of the beam volume up to a considered angular radius. $^{**}$ Typically for 1.25~mm the gain correction is about 8~\% at an elevation of 30~deg (3~\% at 2.14~mm).}
\label{tab:table_err}
\end{table*}

\begin{figure}[t!]
\begin{center}
\includegraphics[width=7cm]{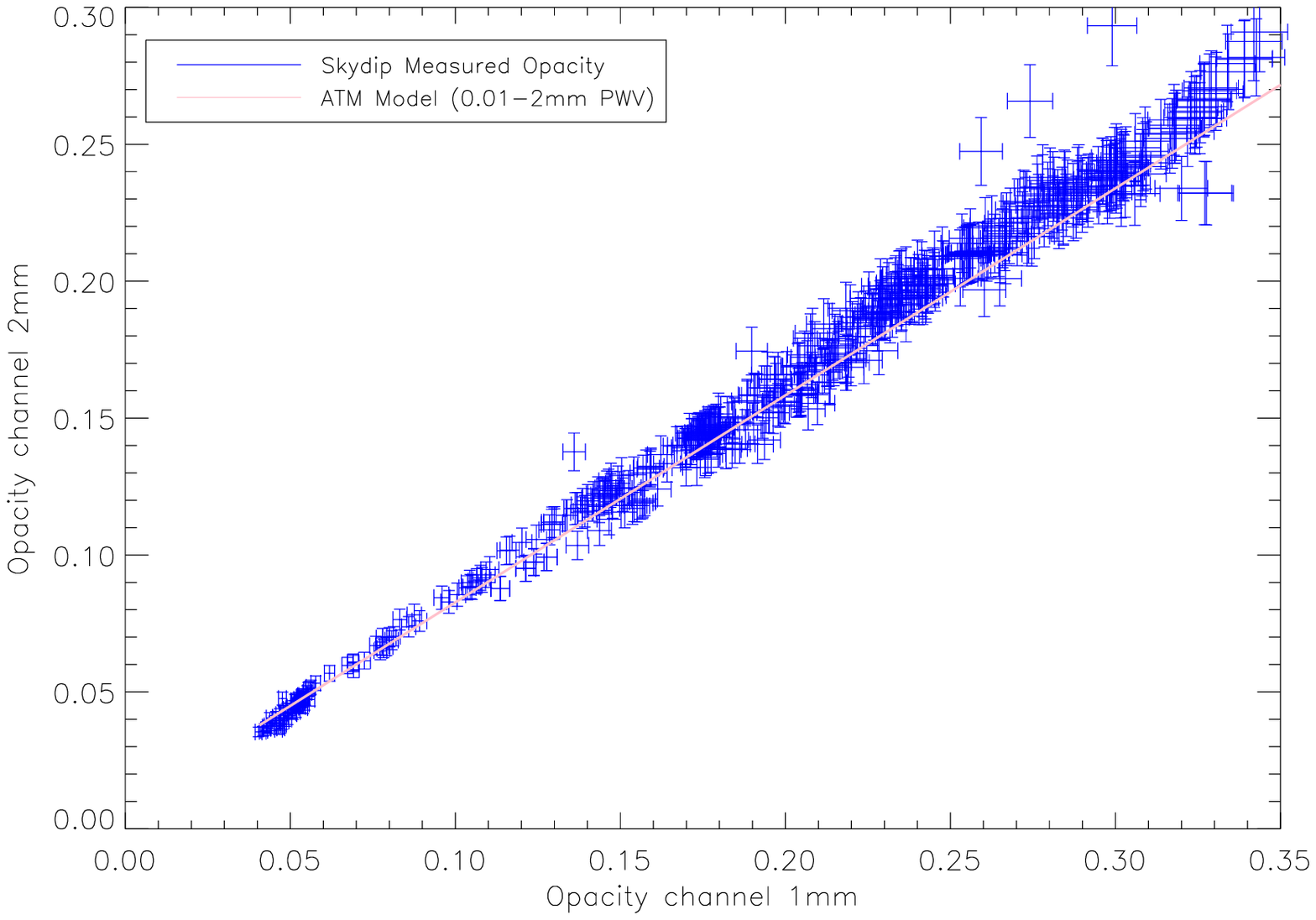} \\
\includegraphics[width=7cm]{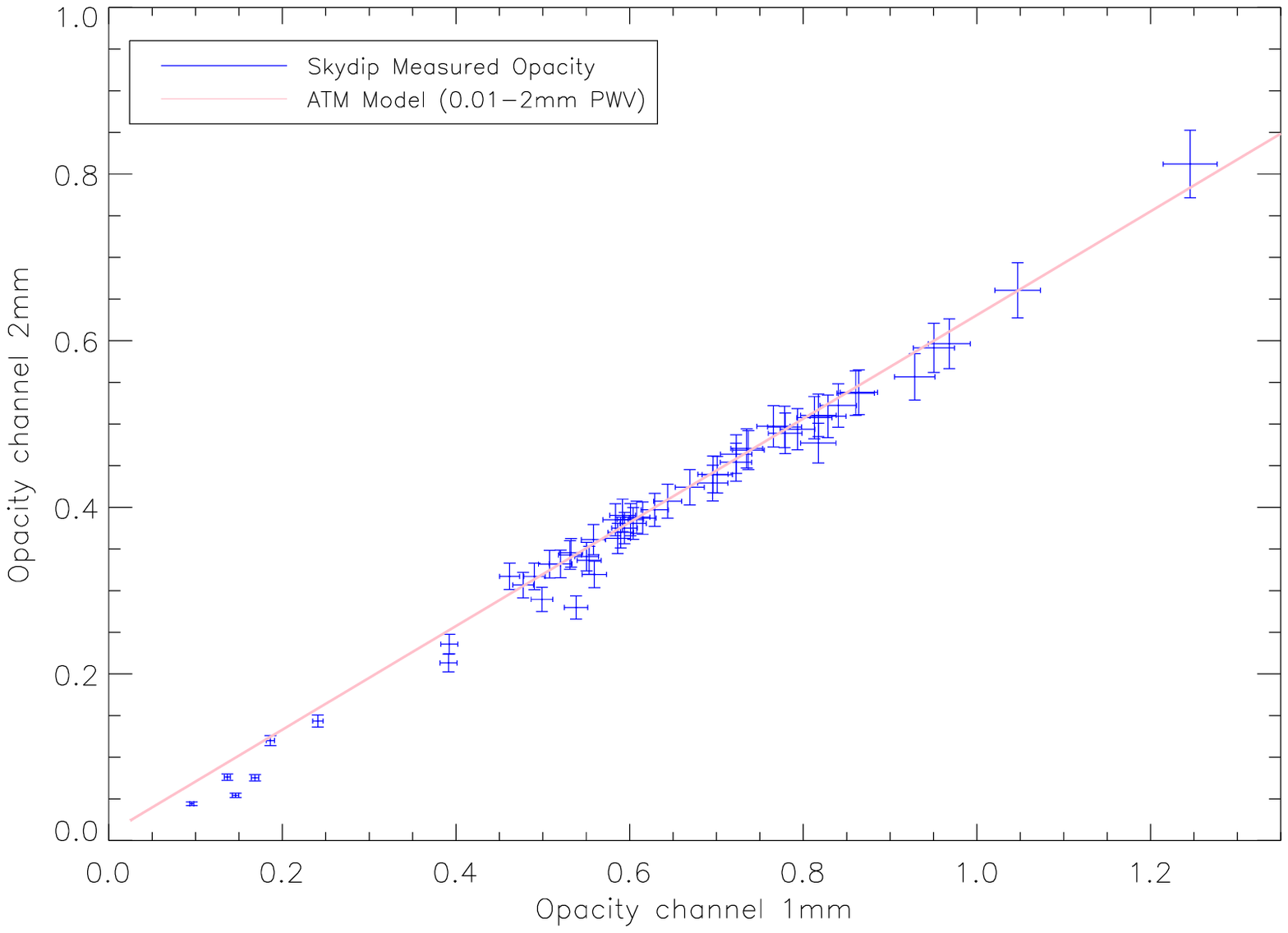}
\end{center}
\caption{Comparison between the atmospheric opacities measured with the \Skydip\ technique
  (blue points) and the ATM model (purple lines). The top plot is for the 2012
  campaign, the bottom plot for the 2013 campaign.}
\label{fig:skydipvsmodel}
\end{figure}


The $F_0$, $C$ coefficients only depend on the response of the detectors.
Since the non-linearities of the KID frequency signal are negligible in the
considered range of backgrounds, the coefficients can be applied to all the
observing campaign to recover the opacity of the considered scan. This is
obtained by inverting Eq \ref{eq:skydip} as

\begin{equation}\label{eq:skydip2}
\tau_{scan_i}=\frac{1}{x_{scan}}\ln{(1-\frac{F^{Ground}_{scan_i}-F_{0_i}}{C_iT_{atm}})}.
\end{equation}
where $\tau_{scan_i}$ is the opacity of the considered scan as measured by a
given detector, $x_{scan}$ corresponds to the air mass at the elevation of the
considered scan, $F^{Ground}_{scan_i}$ is the measured absolute value of the
detector resonance frequency during the scan, and $F_{0_i}$ and $C_i$ are the
coefficients derived from the \Skydip\ technique. The opacity $\tau_{scan}$ is
deduced by averaging $\tau_{scan_i}$ for all valid detectors at a given
wavelength. The brightness of the observed scan map $S^{Ground}$ can be
rescaled onto the scale that would be obtained using detectors outside the
atmosphere ($S^{Star}$) with

\begin{equation}
S^{Star} =  S^{Ground} \cdot e^{ x \tau_{scan}}.
\end{equation}

A initial advantage of this method is that we do not need to perform the \Skydip\ at
the exact time of the source observations to properly correct the atmospheric
contribution in the considered scan. A second advantage is that we can
estimate the atmospheric opacity at the same (azimuth-elevation) position of the source,
instead of the average sky opacity. Finally, with this method of correcting for the atmospheric 
absorption, the \NIKA\ bandpasses are directly taken into account. This method is only limited by the
validity of the air-mass scaling law with elevation (secant model) and by the
degeneracy of the atmospheric temperature with the opacity in our model. To avoid an excessive impact of this degeneracy, we performed a few \Skydip\
procedures per day of observation.

In Fig.~\ref{fig:op} we present the measured opacities for all the mapping scans of the
\NIKA\ observing 2012 (top) and 2013 (bottom) campaigns. During the 2012 campaign
on 22 and 23 of November, the opacity was less then 0.05 for the 1~mm
channel. During the June 2013 campaign, the weather was worse, only permitting
proper observations with the 1.25~mm and 2.14~mm channels for about a few
hours at the end when the opacity was about 0.1 for the 1.25~mm channel.

\subsubsection{Consistency with models}

The consistency of the \Skydip\ technique can be validated by using the ATM
model (\cite{2001IEEE....49.1683C}). We derived the expected opacities integrated into the actual \NIKA\
bandpasses over a range between 0.04 and 20~mm of precipitable water vapor.
In Fig \ref{fig:skydipvsmodel} we present the comparison between the opacities
derived for 1.25~mm and 2.14~mm channels and the ATM model. The plots at the top present the results for the 2012
observing campaign and the bottom plots results for 2013 observing campaign. The
agreement between the measured opacities and the model is good for both
campaigns.


%

%% file: 07_nefd.tex
\begin{figure}[t]
\begin{center}
\includegraphics[scale=0.5,angle=0]{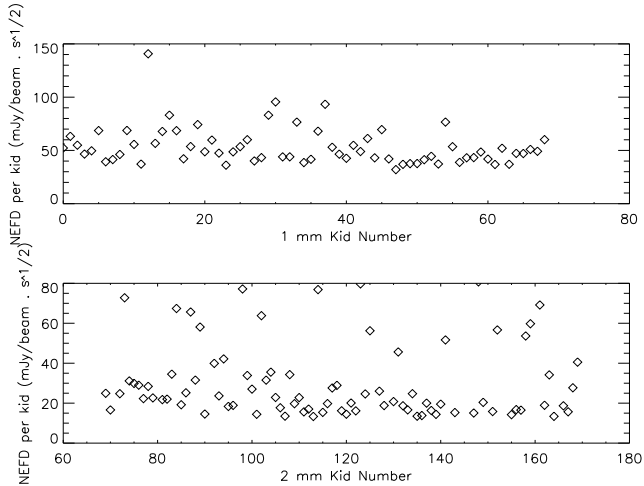}
\caption{NEFD distribution for the two arrays (1.25 mm
  top, 2.14~mm bottom) during the  2012 observation campaign, for the MM18423 scans. The detector
  sensitivity is presented as a function of the KID number, sorted
  according to the resonance frequency. The average NEFD of the two arrays are
  48 and 23~mJy.s$^{1/2}$ at 1.25 and 2.14~mm, respectively, for these
  observations. If we consider only the best 20\% detectors of each array,
  then we get an average NEFD of 39 and 15~mJy.s$^{1/2}$.}
\label{fig:NEFDrun5}
\end{center}
\end{figure}


The flux of a point source is measured via a 2D Gaussian fit in the final average map with a fixed
width and a fixed position. The fit is linear and
involves a constant background, evaluated in a square area with a size of
ten times the FWHM of the Gaussian distribution, in our case 13~arcsec
(resp. 18~arcsec) at 1.25~mm (resp. 2.14~mm). A photometric correction (of
the order of 5~\%) is applied to the flux measurement to account for the
filtering effect induced by the TOI processing.

The noise equivalent flux density (NEFD) is computed as the array-averaged
sensitivity to point sources {\it i.e.} the flux rms obtained in one second of
integration. The average is done via the reciprocal of the square of the
NEFD. It takes the time effectively spent by the array on
the source into account.\\
Figure~\ref{fig:NEFDrun5} presents the NEFD for each valid KID of the array at
1.25~mm (top) and 2.14~mm (bottom) obtained during the observation of
MM18423+5938.  In this case, the average sensitivities on the sky are $\sim
48$ and $\sim 23$ mJy.s$^{1/2}$ at 1.25 and 2.14~mm, respectively. We note that
the sensitivity distribution variation across the array is quite large with the majority of detectors concentrated at the better sensitivity end. If we consider only the best 20~\% detectors of each array,
  then we get an average NEFD of 39 and 15~mJy.s$^{1/2}$.

\begin{figure}[t]
\begin{center}
\includegraphics[scale=0.45]{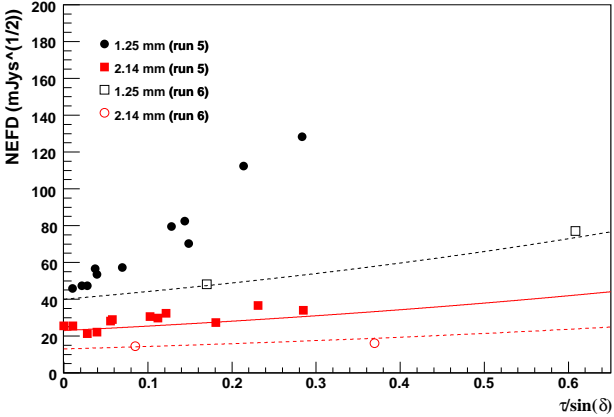}
\caption{Array-averaged NEFD as a function of the line-of-sight opacity $\tau/
sin(\delta)$. Black dots (resp. red squares) refer to the 1.25~mm array (resp. 2.14~mm) for the 2012 observation campaign, during which 
the trend at  2.14~mm is compatible with a $exp(\tau/sin(\delta))$ behavior (as shown with the red curve),
 while a departure from this behavior is observed at  1.25~mm. The latter can be attributed to the  degradation of the KID resonance quality
  factor with increasing power load. Open circles (resp. open squares) refer to the 1.25~mm array (resp. 2.14~mm) for the 2013 observation campaign. Very few points have been measured,  but the achieved sensivities are much better and do follow an expected exponential 
  trend.}
\label{fig:NEFDfull}
\end{center}
\end{figure}

Figure \ref{fig:NEFDfull} presents the array-averaged NEFD of the \NIKA\ camera
at 1.25 and 2.14~mm as a function of the line-of-sight opacity $\tau/
sin(\delta)$, where $\tau$ is the zenith opacity and $\delta$ the
elevation. The NEFD measurements are  obtained during the
observations described in Table~\ref{tab:table_sed}.  At zero zenith opacity,
the array-averaged NEFD are $\sim 40$ and $\sim 23$ mJy.s$^{1/2}$ at
1.25 and 2.14~mm, respectively, which give an indication of the performance of the 2012
\NIKA\ camera in optimal weather conditions.  As expected, the sensitivities
degrade with increasing opacity, corresponding to worse weather conditions. At
2.14~mm, we observe, as expected, a degradation due to the opacity effect, following 
$exp(\tau/sin(\delta))$, while the degradation seems to be
enhanced at 1.25~mm. Nevertheless, the sky noise decorrelation seems to work  in those
cases as well. We therefore confirm that the increased power on kids somehow reduces
their performance.

The sensitivity at 1.25~mm was limited by a saturation effect on the
readout electronics, which has been corrected since. By only using eight
detectors, we have been able to detect  the faint source MM18423. In that
case, the measured, averaged 1.25~mm NEFD is 27~mJy.s$^{1/2}$ to be compared with 48~mJy.s$^{1/2}$ for the full-array observation 
(fig.~\ref{fig:NEFDrun5}). 

SXDF1100.001 has been observed twice during this observation campaign, with two different integration times 
(68 and 258 min), and we
used it to have a first estimate of the evolution of NEFD and noise with integration time. 
We note that the noise integrates down because the NEFD is stable, and the noise is divided by a 
factor 2 when increasing the integration time by a factor 4.

The 2013  observation campaign allowed us to test new arrays. Although an additional filter
has limited the 1.25~mm band efficiency, we were able to measure the 
array-averaged NEFD for two sources (HFLS3 and MM18423) in weather conditions far from 
optimal ($\tau$ up to 0.6).  As shown in figure \ref{fig:NEFDfull}, the achieved sensivities are much better (about 1/3 lower) 
and do follow an expected exponential trend. We extrapolate a sensitivity 
at zero zenith opacity of $\sim 40$ mJy.s$^{1/2}$ at 1.25~mm and $\sim 14$ at 2.14~mm. This improvement is related to technological parameters optimization 
as explained in section \ref{dac}.

In conclusion, the \NIKA\ instrument has shown an array-averaged NEFD on the sky for point sources of 
40 and 14~mJy.s$^{1/2}$ at 1.25 and 2.14~mm in optimal weather conditions. If we consider typical NIKA observing conditions (zenith opacity = 0.1, air mass = 1.5) the array-averaged NEFD results 46~mJy.s$^{1/2}$ at 1.25~mm channel and 15~mJy.s$^{1/2}$ at 2.14~mm channel. 
Hower, we expect the 1.25~mm channel sensitivity to be improved in the next  observation campaigns.

\begin{table*}
\begin{center}
\begin{tabular}{lllrrrrr}
\hline
\hline
Source name & RA & Dec & $F_{\nu}\mathrm{(1.25~mm)}$ & $F_{\nu}\mathrm{(2.14~mm)}$       & $T_{int} (min)$ &  $\tau_{1}$ & $\tau_{2}$ \\
\hline
            & 2000 & 2000 &mJy          &mJy              & min      &              & \\
\hline
HLS091828     &  09:18:28.600  &  +51:42:23.300 & $36.7   \pm  4.6$  & $8.3 \pm 0.7 $   & 83 & 0.27 & 0.22 \\
MM18423       &  18:42:22.500  &  +59:38:30.000 & $33.6   \pm  3$  & $6.3 \pm 0.8 $   & 37 & 0.02 & 0.03 \\
SXDF          &  02:18:30.600  &  $-$05:31:30.000 & $28   \pm  1.5$  & $4.1 \pm 0.4 $   & 258& 0.03 & 0.03 \\
HFLS3$^{r6}$   &  17:06:47.800  &  +58:46:23.000 & $16   \pm  2$  & $4.0 \pm 0.6 $  & 16 & 0.14 & 0.07 \\ \hline
Arp220        &  15:34:57.100  &  +23:30:11.000 & $243  \pm  3 $  & $52.8  \pm 0.8 $   & 47 & 0.11 & 0.09 \\
HAT084933     &  08:49:33.400  &  +02:14:43.000 & $13   \pm  3$  & $1.3 \pm 0.8 $   & 59 & 0.08 & 0.07 \\
HAT133008     &  13:30:08.560  &  +24:58:58.300 & $16   \pm  3$  & $4.5 \pm 0.8 $   & 55 & 0.14 & 0.10 \\
PSS2322+1944  &  23:22:07.200  &  +19:44:23.000 & $<4.6         $  & $<1.7          $   & 57 & 0.06 & 0.05 \\
GRB121123A    &  20:29:16.290  &  $-$11:51:35.900 & $<15        $  & $<1.7          $   & 69 & 0.22 & 0.18 \\
4C05.19       &  04:14:37.800  &  +05:34:42.000 & $<6.2        $   & $26.3 \pm 0.8          $   & 15 & 0.13 & 0.11 \\
ZZTauIRS      &  04:30:51.714  &  +24:41:47.510 & $77  \pm  2$   & $16.2  \pm 0.8 $   & 40 & 0.01 & 0.00 \\
CXTau         &  04:14:47.865  &  +26:28:11.010 & $<4.6      $   & $<1.7          $   & 40 & 0.02 & 0.01 \\
\hline \hline

\end{tabular}
\end{center}
\caption{\NIKA\ Flux and sensitivity of a collection of point sources. 
The integration time is given in minutes. 
The average zenith opacities are given in the last two columns for the two
\NIKA\ wavelengths. 
Upper limits are given as $2~\sigma$. 
Most of the data were taken from 19 to 24 November 2012. 
}
\label{tab:table_sed}
\end{table*}


%% file: 08_skyobs.tex
\begin{figure*}[t]
\begin{center}
\includegraphics[scale=0.35,angle=0]{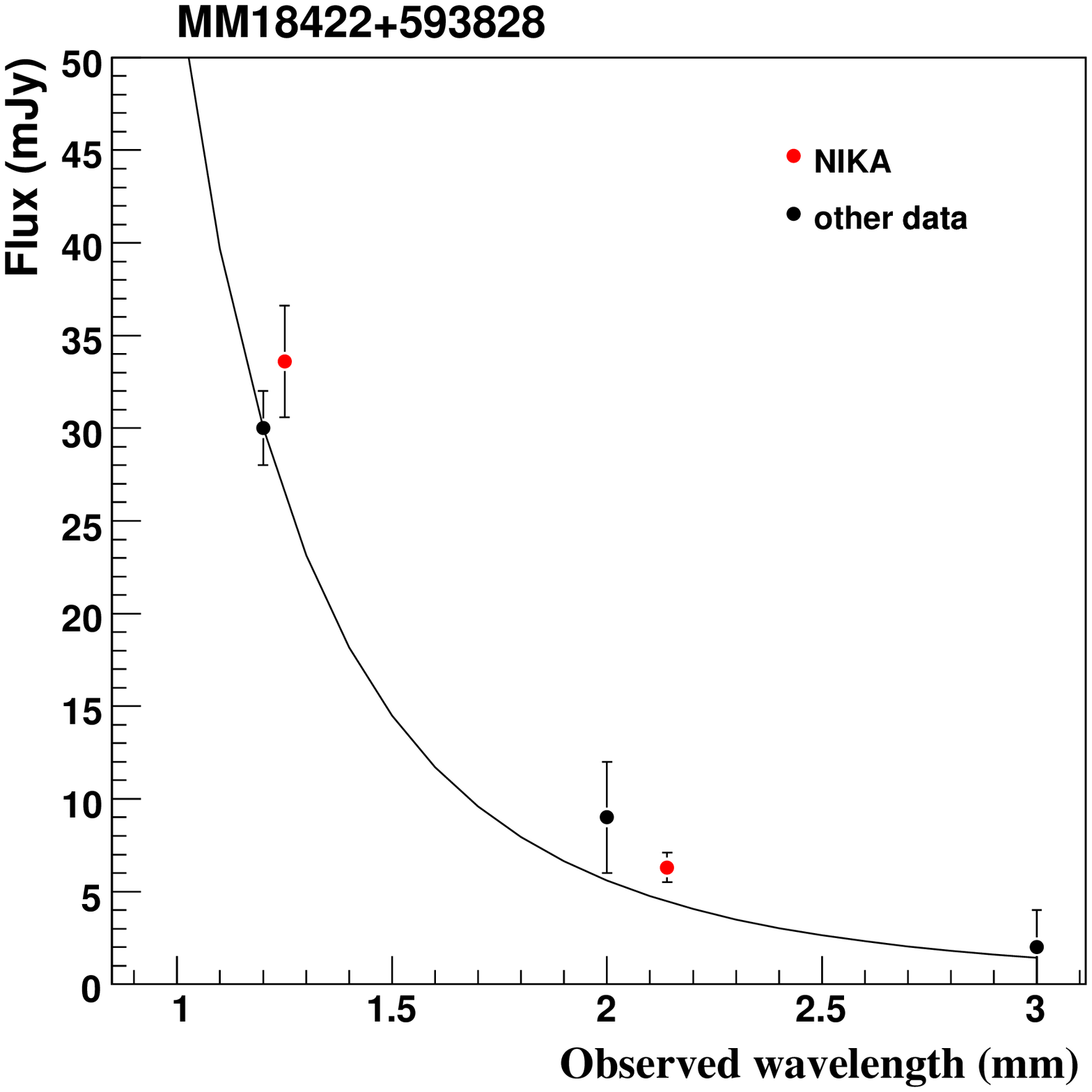}
\includegraphics[scale=0.35,angle=0]{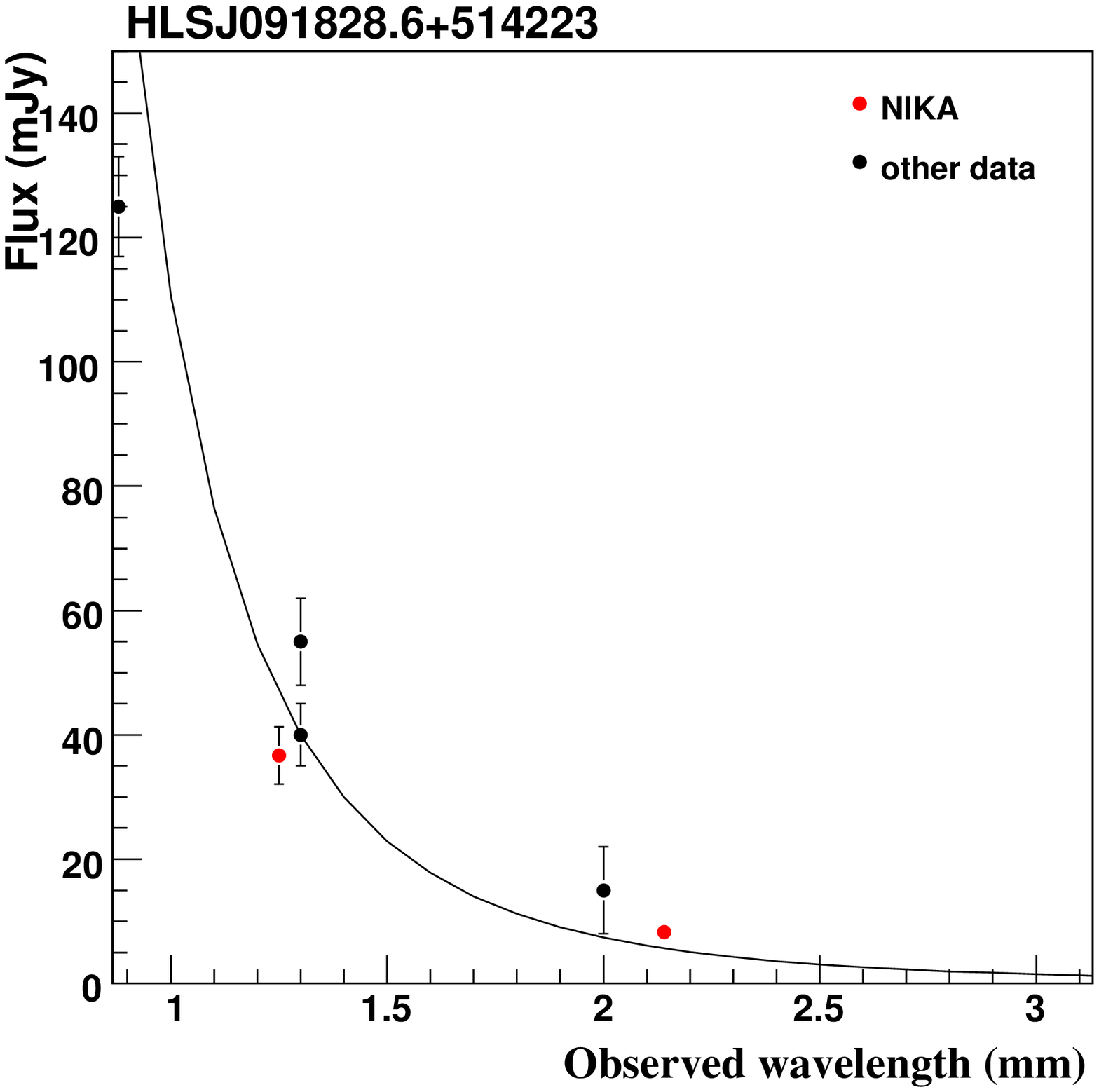}
\includegraphics[scale=0.35,angle=0]{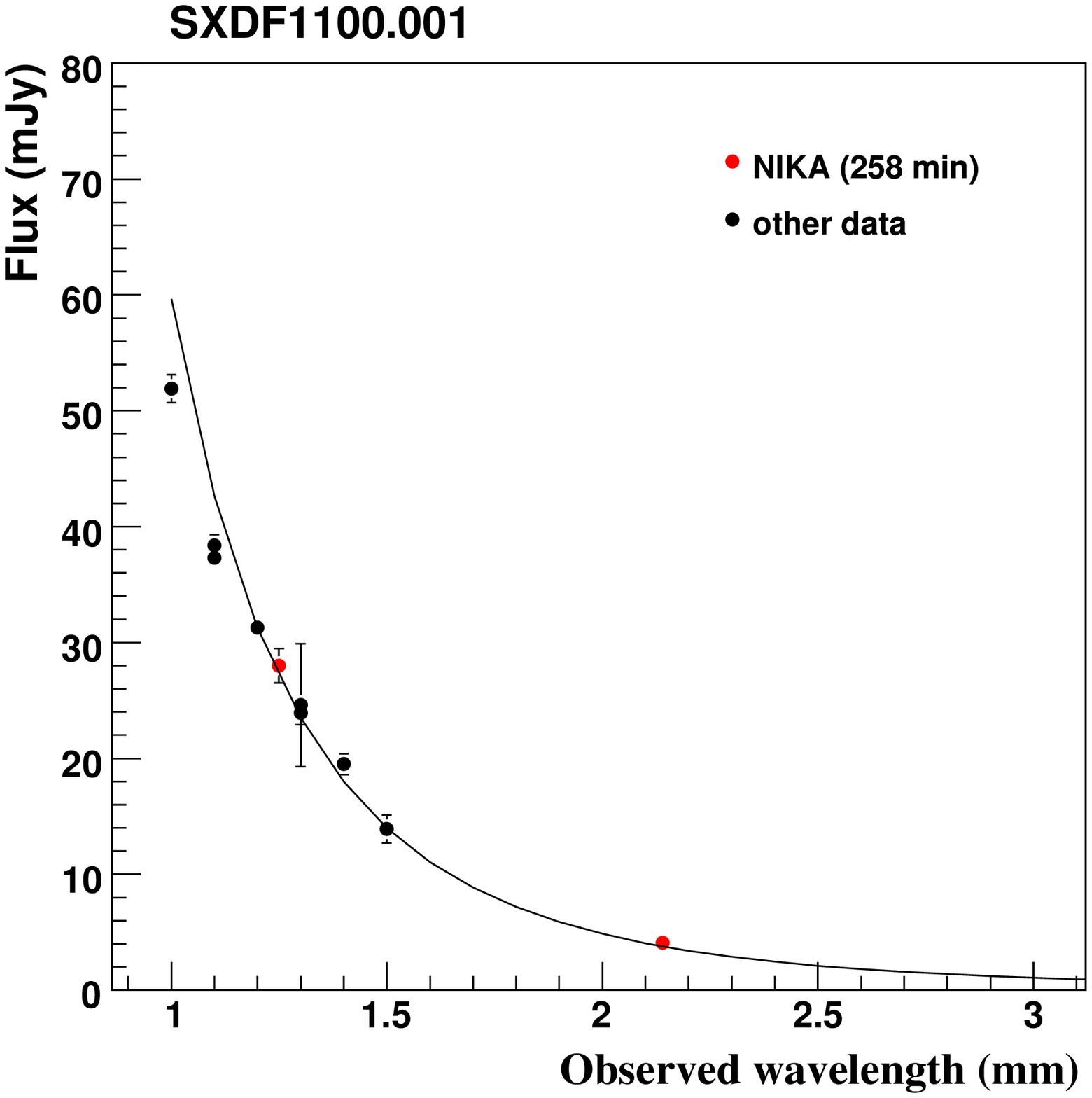}
\includegraphics[scale=0.35,angle=0]{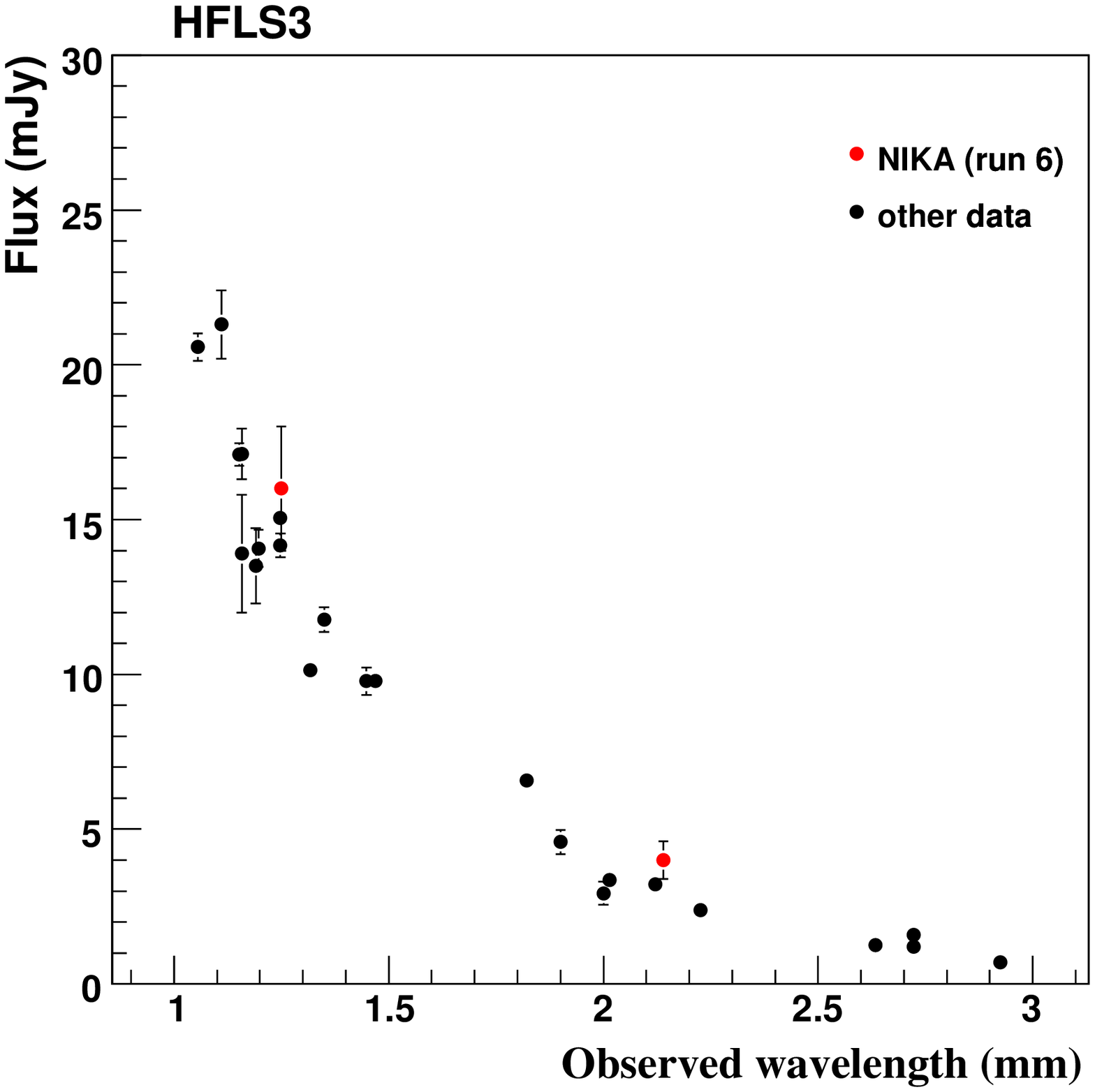}
\caption{SED of selected pointlike sources observed with \NIKA\ 2012 and 2013 observation campaigns, 
flux (mJy) as a function of the observed wavelength (mm). 
Upper left : MM18423+5938 observed by \NIKA\ 2012 observation campaign (red points) and previous millimeter observations 
(black points) (\cite{Lestrade:2010wm}). 
Solid line indicates a modified grey body with $\beta=1.5$ and $\rm T=24  \ {\rm K}$, see (\cite{McKean:2011nk}). 
Upper right :  HLSJ091828.6+514223 observed by \NIKA\ 2012 observation campaign (red points) and previous millimeter 
observations (black points), see (\cite{2012A&A...538L...4C}) and references therein. 
Solid line indicates a modified grey body with $\beta=2$ and $\rm T=52  \ {\rm K}$, see 
\citep{2012A&A...538L...4C}.
Lower left : SXDF1100.001 observed by \NIKA\ 2012 observation campaign (red points) and previous millimeter observations  
(black points), see (\cite{Ikarashi:2010ar}) and references therein. 
Solid line indicates a modified grey body with $\beta=1.9$ and $\rm T=20  \ {\rm K}$, see 
\citep{Ikarashi:2010ar}.
Lower right : HFLS3 observed by \NIKA\ 2012 observsation campaign (red points) and previous millimeter observations  (black points), see (\cite{2013Natur.496..329R}) and references therein.}
\label{fig:sedpointlikesources}
\end{center}
\end{figure*}

\begin{figure*}[t!]
\begin{center}
\includegraphics[height=5cm,width=6cm]{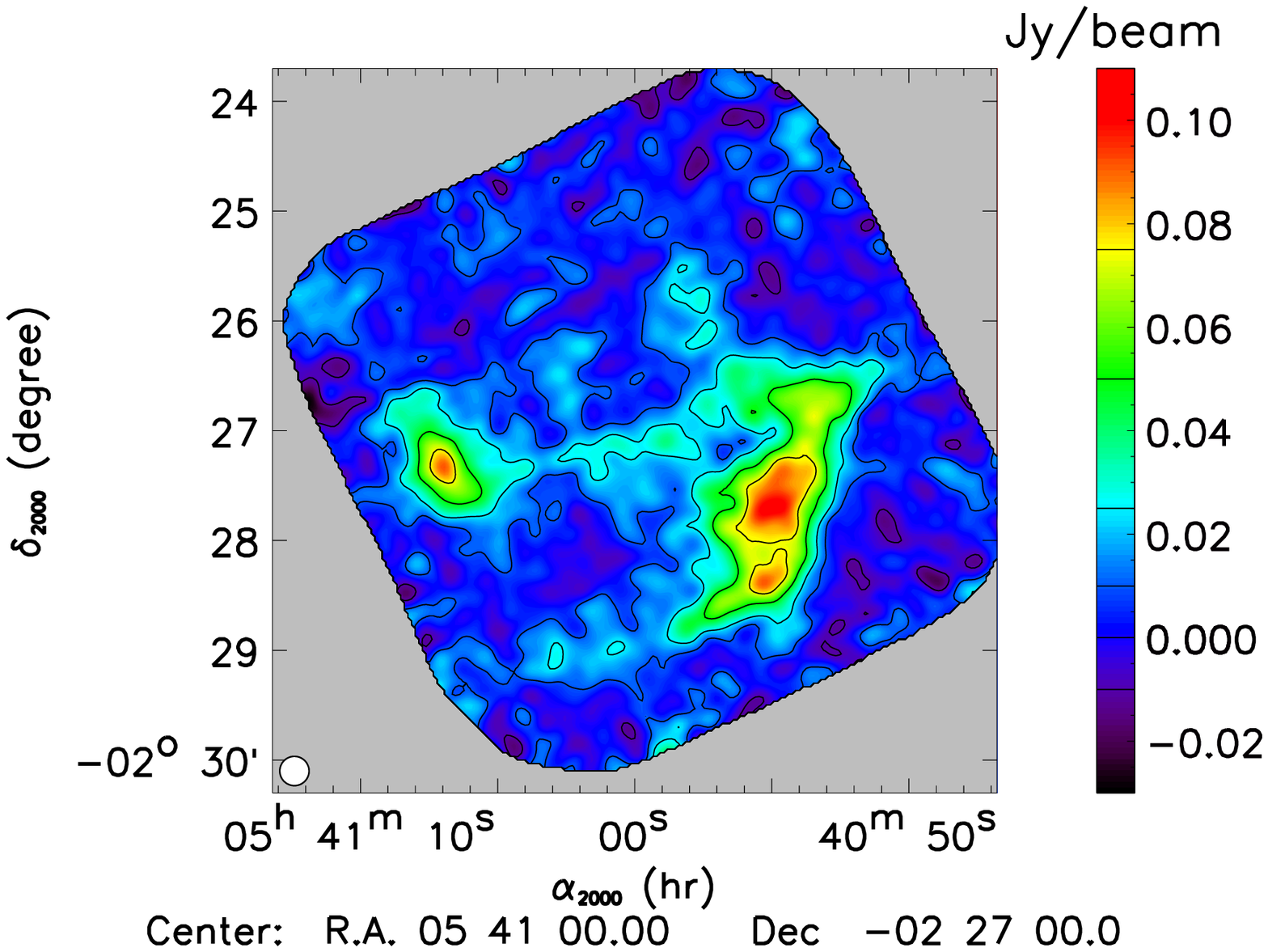} 
\includegraphics[height=5cm,width=6cm]{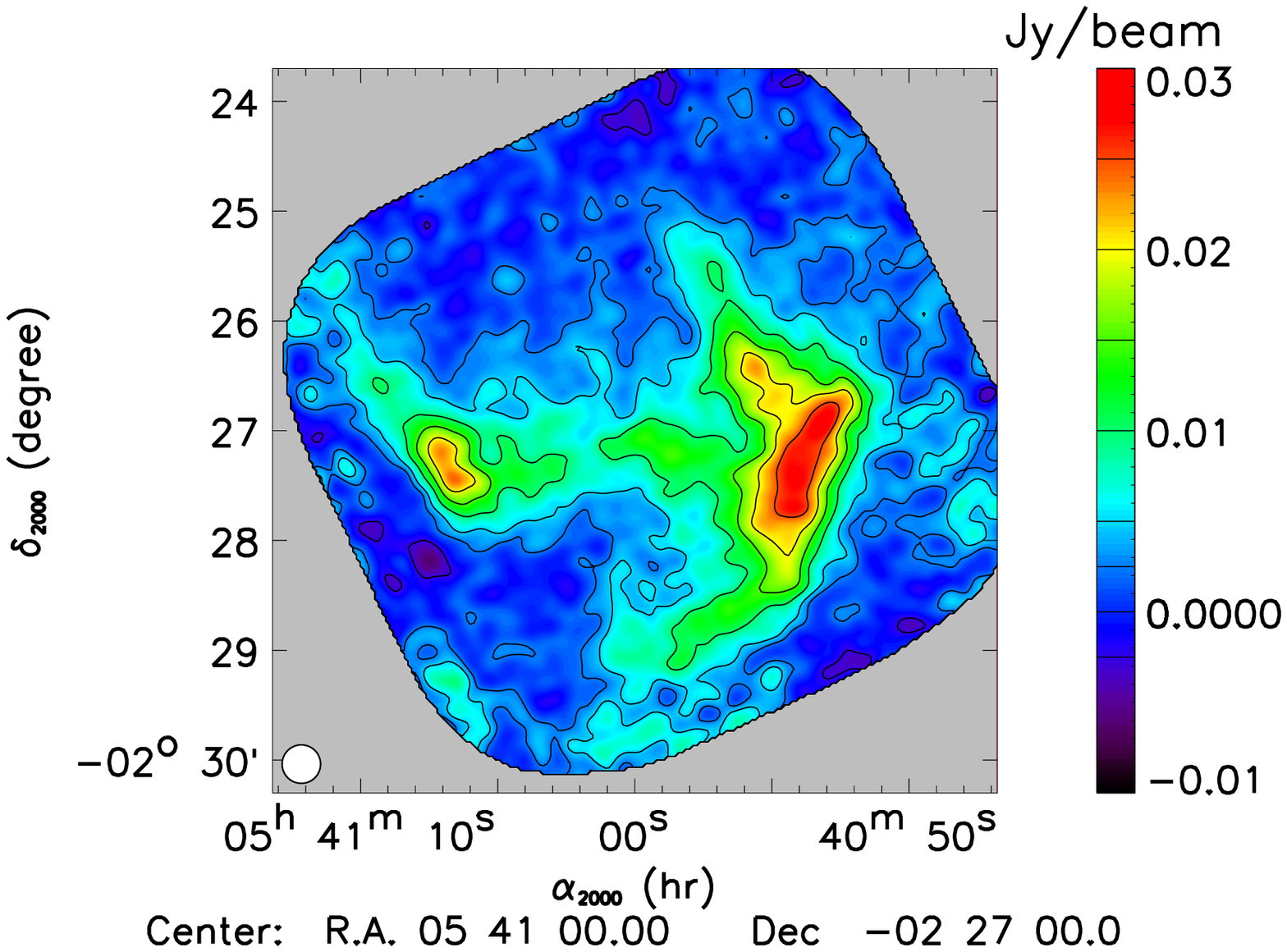}
\includegraphics[height=5cm,width=6cm]{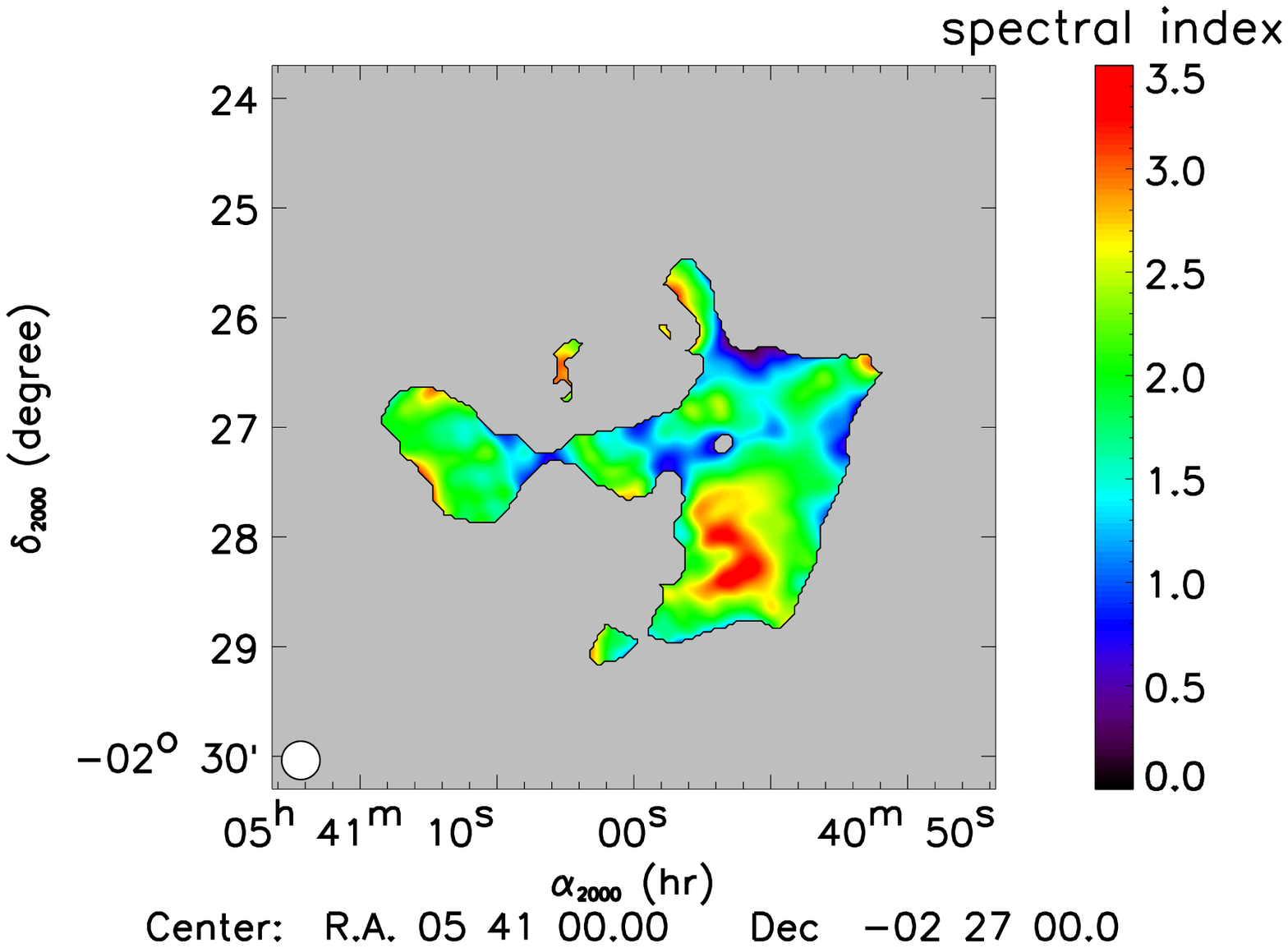}
\includegraphics[width=6cm]{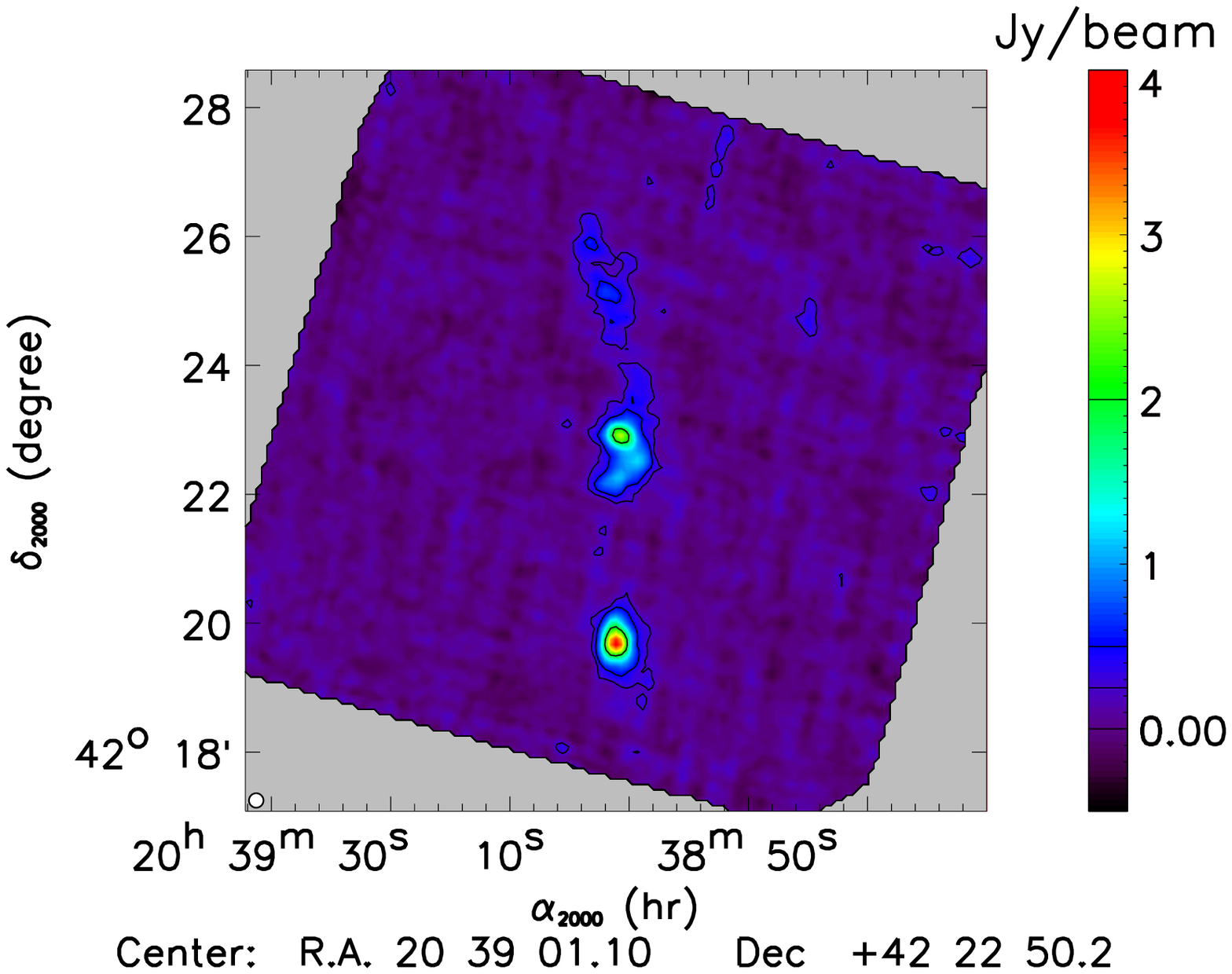} 
\includegraphics[width=6cm]{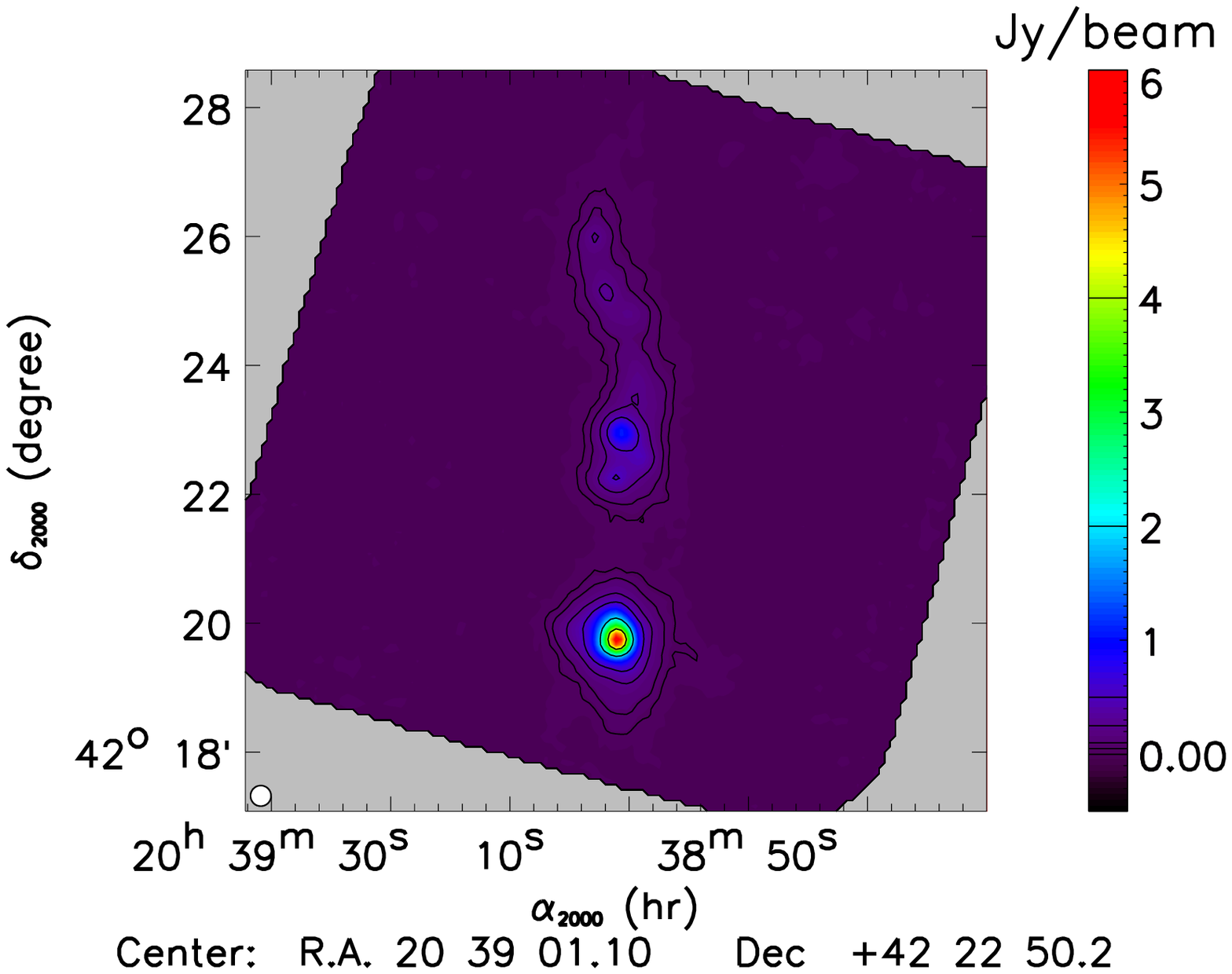}
\includegraphics[width=6cm]{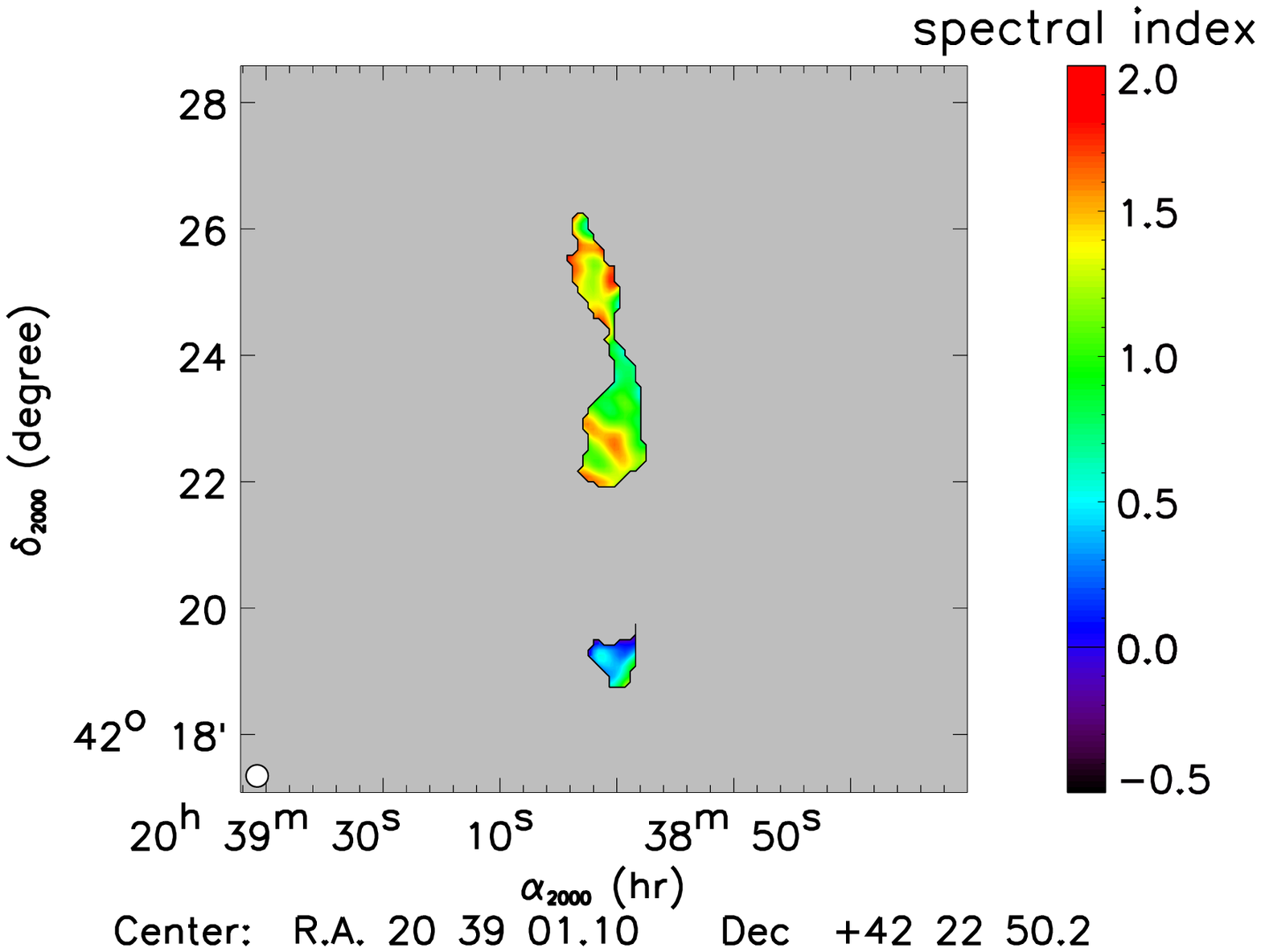}
\end{center}
  \caption{Examples of 1.25 (left) and 2.14~mm (middle) NIKA maps of well-known extended sources. From top to bottom we present
the Horsehead nebula and the DR21 OH complex. 
Spectral index maps are presented in the right column.}
\label{fig:extended_sources}
   \end{figure*} 
 
 \begin{table*}
\begin{center}
\begin{tabular}{ccccccccc}
\hline
\hline
Source & RA & DEC  & Integration time   & Noise rms 1.25 mm   &  Noise rms 2.14 mm   & $\tau_{1}$ & $\tau_{2}$ \\
\hline
&  [deg] & [deg] &  [hours] &   [mJy/beam]  &   [mJy/beam]  &  & \\
\hline
Horsehead  &  85.25     &    -2.45  &   1.57 & 14.5  &  2.0   &   0.027 &   0.022 \\
DR 21 OH     &  309.75   &   16.27 &   0.54 & 181.0& 12.6 & 0.27 &  0.22 \\
\hline \hline
\end{tabular}
\end{center}
\caption{Main properties of the maps of extended sources: center of the map, integration time, rms noise, and opacity
at the two observation frequencies.}
\label{tab:extended_sources}
\end{table*}

The \NIKA\ camera was used during the November 2012 and June 2013
campaigns to observe point like sources, in order to assess the \NIKA\
photometry, and extended sources to demonstrate the possibility to reconstruct
angular scales up to few arcminute.

\subsection{Millimeter spectral energy distribution (SED) of selected point sources}
\label{SED}
Point-like sources were observed during the good weather conditions in
the November 2012 and June 2013 campaigns. They were selected to be faint
but detectable, in order to assess the \NIKA\ photometry
with fluxes of a few tens of mJy.
Table~\ref{tab:table_sed} describes the measured point-source fluxes at 1.25
and 2.14~mm. Flux errors are statistical only. The opacity corresponds to a
value averaged over the scans.  Figure~\ref{fig:sedpointlikesources} presents
the spectral energy distribution (SED) of a selection of four sources observed
in the 2012 and 2013 campaigns and compares it with previous measurements. We note
that \NIKA\ observations are in good agreement with previous
observations. We have chosen to observe the following point-source high redshift submillimeter
galaxies (SMG): MM18423+5938, HLSJ091828.6+514223, SXDF1100.001, and HFLS3. For the two campaigns, the weakest source detected
was HFLS3 with a flux of $16\pm 2 \ {\rm mJy}$ at 1.25 mm and $4.0 \pm 0.6 \ {\rm mJy}$ at 2.14~mm.

\subsubsection*{MM18423+5938}
MM18423+5938 is a submillimeter galaxy at $z = 3.93$ discovered
serendipitously (\cite{Lestrade:2009ef}) in a search for cold debris disks
around M dwarfs with MAMBO-2 at the IRAM 30-m millimeter telescope.
Flux densities at 1.2~mm, 2~mm, and 3~mm were measured
(\cite{Lestrade:2010wm}.  The relatively high flux (about 30~mJy at 1~mm) is
explained by the fact that this SMG is gravitationally lensed, as can be
assessed by the observation of the CO emission, which is consistent with a
complete Einstein ring with a major axis diameter of $1.4^{\prime\prime}$
(\cite{Lestrade:2011qq}.  
Figure~\ref{fig:sedpointlikesources} (upper left) presents the SED of
MM18423+5938 with NIKA measurements and previous ones \citep{Lestrade:2010wm}. J.~P.~McKean {\it et al.}  have fitted the SED with a
single temperature modified blackbody spectrum, with $\beta=1.5\pm 0.5$ and
$\rm T=24^{+7}_{-5} \ {\rm K}$, although the lack of measurements at high
frequencies allows for a wider range of temperatures (\cite{McKean:2011nk}. 
Since NIKA data are not included in the fit, we use it to assess the NIKA photometry. It is presented as a solid line 
in figure~\ref{fig:sedpointlikesources} (upper left).


\subsubsection*{HLSJ091828.6+514223} 
HLSJ091828.6+514223 is an exceptionnally bright source at millimeter and submillimeter wavelength, discovered
behind the z=0.22 cluster Abell 773 (\cite{2012A&A...538L...4C}). It appears to
be a strongly lensed submillimeter galaxy (SMG) at a redshift of z=5.24, the
lens being an optical source lying in the neighborhood. It has been measured
at 2\,mm (IRAM-30\,m EMIR), 1.3\,mm (SMA), and 0.88 mm (SMA) (see
\cite{2012A&A...538L...4C} and references
therein). Figure~\ref{fig:sedpointlikesources} (upper right) presents the SED
of HLSJ091828.6+514223 with \NIKA\ measurements and previous ones \citep{2012A&A...538L...4C}).
F. Combes {\it et al.}  have fitted the SED with a
single temperature modified blackbody spectrum with $\beta=2$ and
$\rm T=52 \ {\rm K}$.
As NIKA data are not included in the fit, we use it to assess the NIKA photometry. The SED is presented as a solid line 
in figure~\ref{fig:sedpointlikesources} (upper right).

\subsubsection*{SXDF1100.001}
SXDF1100.001 (also known as Orochi) is an extremely bright (50 mJy at 1 mm)
submillimeter galaxy, discovered in 1.1\,mm observations of the
Subaru/XMM-Newton Deep Field using AzTEC on ASTE (\cite{Ikarashi:2010ar}).  It
is believed to be a lensed, optically dark SMG lying at z=3.4 behind a
foreground, optically visible (but red) galaxy at z=1.4.\\
Continuum flux densities at millimeter wavelengths have been measured by SMA,
Carma, Z-SPEC/CSO, and AzTEC/ASTE (see \cite{Ikarashi:2010ar} and references
therein). Figure~\ref{fig:sedpointlikesources} (lower left) presents the SED
of SXDF1100.001 with \NIKA\ measurements. To our knowledge, this is the first
measurement at    2\,mm.
S. Ikarashi {\it et al.}  have fitted the SED with a
single-temperature, modified blackbody spectrum, with $\beta=1.9$ and
$\rm T=20 \ {\rm K}$ \citep{Ikarashi:2010ar}.
As NIKA data are not included in the fit, we use it to assess the NIKA photometry. It is presented as a solid line 
in figure~\ref{fig:sedpointlikesources} (lower right).

\subsubsection*{HFLS3}
HFLS3 has been reported as a massive starburst galaxy at redshift 6.34
(\cite{2013Natur.496..329R}).  Continuum emission has been measured over a broad
wavelength range, in particular at millimeter wavelengths (Z-Spec, PdBI/IRAM,
CARMA/Caltech, Gismo/IRAM-30m) (see \cite{2013Natur.496..329R} and references
therein). Figure~\ref{fig:sedpointlikesources} (lower right) presents the SED
of HFLS3 with \NIKA\ measurements, which are in good agreement.

\subsubsection*{Other observations} 
These four comparisons of \NIKA\ observations at 1 and 2~mm with previous
observations allow us to assess the NIKA photometry.  Other observations
(shown in the lower part of Table~\ref{tab:table_sed}) allow probing the
diversity of sources and fluxes and measuring the camera performance in various
background conditions. In particular,  Arp220 (aka IC 4553) is the closest (z = 0.018) ultraluminous
infrared galaxy, known to be a merger system composed of a double nucleus. 
Continuum flux densities at millimeter wavelengths has been measured \citep{Sakamoto, Scoville, woody} 
but particular attention must be paid to line contamination due to the prolific molecular line emission. 
S.~Martin {\it et al.} have shown  that 
the contamination of molecular emission to the flux 
constitutes 28~\% of the overall flux between 203 and
241~GHz  \citep{Martin:2010gg}.\\  
HAT084933, HAT133008, PSS2322+1944, and 4C05.19 are high-redshift
point sources. GRB121123A is a gamma-ray burst that happened during the 2012
campaign. ZZTauIRS and CXTau are stars detected by Herschel.

 \begin{table*}[t!]
\begin{center}
\begin{tabular}{ccc}
\hline
\hline
{\bfseries Array} & 1.25~mm & 2.14~mm  \\
\hline \hline
{\bfseries Valid pixels} & 80~(run5)~190~(run6) & 100~(run5)~125~(run6) \\
{\bfseries Field of view}~(arcmin) & 2.2 & 2.2 \\
{\bfseries Band-pass}~(GHz) & 125-175   & 200-280~(run5)~200-270~(run6)  \\
{\bfseries FWHM}~(arcsec) &  12.3  & 18.1 \\
{\bfseries Sensitivity}~(mJy$\cdot s^\frac{1}{2}$) & 40 & 14  \\
{\bfseries Mapping speed}~(arcmin$^2$/mJy$^2$/hour) & 8 & 57  \\
\hline \hline
\end{tabular}
\end{center}
\caption{Summary of the NIKA characteristics and performance. Observation campaign 2012 is referred to as run5. Observation campaign 2013 is referred to as run6}
\label{tab:table_fin}
\end{table*}

\subsection{Mapping of extended sources}   
\label{ES}

During the 2012 and 2013 observation campaigns we observed several well-known extended sources simultaneously at 1 and 2~mm to test the capabilities of
the \NIKA\ camera to recover large angular scales up to few arcmin. We performed
both elevation and azimuth raster scans to ensure an homogeneous coverage of
the mapped area. The size of the scans and integration time have been adapted
to each source. We present two examples of extended sources, namely the
Horsehead Nebula and DR21 OH regions that have been observed
during the 2012 campaign. The center pointing position, the size of the scans,
and the integration time for each sources are given in
Table~\ref{tab:extended_sources}. We also present the median rms of the map
and atmospheric opacity for the 1.25 and 2.14~mm maps.

\subsubsection*{The Horsehead nebula}
The Horsehead nebula is a dark protrusion that emerges from the L1630 cloud in the Orion B molecular complex at about 400~pc. This condensation is illuminated by the 09.5~V star $\sigma$Ori which is at a distance of 0.5$^{\circ}$ from the cloud. It presents a photon-dominated region (PDR) on its western side, which is seen edge-on  \citep{2003A&A...410..577A}. 

We concentrate here on the outer neck \citep{2005A&A...440..909H}, which
consists of the PDR, the nose, the mane and the jaw. In the top row of
Figure~\ref{fig:extended_sources}, we show the 1.25 (left) and 2.14 (middle)~mm
NIKA maps. The typical one-sigma error in the map with the pixel size of 3 arcsec is 14.5~mJy/beam at 1.25~mm and 2 at 2.14~mm.  The NIKA 1.25~mm is consistent with the
1.2\,mm continuum map \citep{2005A&A...440..909H} obtained with MAMBO2, the
MPIfR 117-channel bolometer array from 30~m IRAM telescope \citep{1992ESASP.356..207K}. In the NIKA maps we clearly
observe the PDR region whose morphology changes significantly
from one frequency to another. This is also obvious in the spectral index map
presented in the right hand figure that ranges between 2 and 5. The northern part of the PDR presents a
significantly flatter spectral index (about 4.5) than the southern part and
the mane (about 3.5). Two possible explanations are CO 2-1 contamination at 1
mm or high dust emissivity spectral index\citep{2013ApJ}. 




\subsubsection*{DR21 complex}
DR21 is a giant star-forming complex located in the constellation Cygnus
$\sim 3$ kpc from Earth \citep{1982ApJ...261..550C, DR21_2, DR21_1}.  H2O masers
\citep{1977A&AS...30..145G} and a map of the 1.3-mm continuum emission
\citep{2005IAUS..227..151M} show that DR21 belongs to a north-south oriented
chain of massive star forming complexes. DR21 is composed from north to south
of three main regions DR21, DR21 OH, and DR21 Main (DR21M). DR21M has a mass of
$\sim$ 20000 M$\sun$ and contains one of the most energetic star forming
outflows detected to date \citep{1991ApJ...366..474G,1991ApJ...374..540G,1992ApJ...392..602G}.

We present in the bottom row of Figure~\ref{fig:extended_sources} the NIKA
maps of the DR21 complex at 1.25 (left) and 2.14 (right)~mm. We observe in
them the three main regions in the complex, the most intense being DR21OH and
DR21M. As shown in the right hand plot of the row, the spectral characteristics of
these two regions are significantly different. DR21M has a flatter spectrum in
the frequency range of NIKA.  We suspect the presence of strong free-free
emission.

%% file: 09_conclusion.tex
In this paper we have shown that \NIKA\ is a competitive instrument for
millimeter wave astronomy using KID detector technology.
We presented several instrumental and data analysis improvements including:
\begin{itemize}
\item a reliable optimization of the detectors working point that significantly increased the number of valid detectors and their responsivity;
\item an automatic self atmospheric absorption correction along the line of sight;
\item a data analysis pipeline adapted to KID specifics. 
\end{itemize}
These lead to a significant improvement of the performance in terms of measured NFED
and to accurate photometry on point sources.

Table \ref{tab:table_fin} summarizes the main \NIKA\ 
characteristics and performance as measured on the sky. We obtained a 
sensitivity (averaged over all valid detectors) of 40 and 14~mJy.s$^{1/2}$
for the best weather conditions for the 1.25~mm and 2.14~mm arrays, respectively,
estimated on point like sources.
Additionally, the camera performance 
can be quantified with its mapping speed: the area that can be observed per unit time at a given sensitivity.    
\\
\\
The future \NIKAii\ will be made of about 1000 detectors at 2.14~mm and 2 $\times$ 2000 at
1.25~mm with a circular field of view of $\sim 6.5$ arcmin diameter. \NIKAii\
will be commissioned at the end of 2015. In addition the \NIKAii\ instrument
will have linear polarization capabilities at 1.25~mm. The performance in
polarization will be tested in the \NIKA\ camera during the 2014 observation
campaigns.